\documentclass[iop,numberedappendix]{emulateapj}

\slugcomment{ }
\usepackage[breaklinks,colorlinks,urlcolor=blue,citecolor=blue,linkcolor=blue]{hyperref}

\shorttitle{Gravitationally unstable condensations revealed by ALMA}
\shortauthors{Ohashi et al.}

\begin{document}


\title{Gravitationally unstable condensations revealed by ALMA in the TUKH122 prestellar core in the Orion A cloud}


\author{Satoshi Ohashi\altaffilmark{1,2,3}, 
Patricio Sanhueza\altaffilmark{2}, Nami Sakai\altaffilmark{1}, Ryo Kandori\altaffilmark{4},  Minho Choi\altaffilmark{5}, Tomoya Hirota\altaffilmark{2,6}, \\ Quang Nguy$\tilde{\hat{\rm e}}$n-Lu'o'ng\altaffilmark{5}, and Ken'ichi Tatematsu\altaffilmark{2,6}
}

\affil{${}^{1}$\  RIKEN, 2-1, Hirosawa, Wako-shi, Saitama 351-0198, Japan; \href{mailto:satoshi.ohashi@riken.jp}{satoshi.ohashi@riken.jp}}
\affil{${}^{2}$\ National Astronomical Observatory of Japan, 2-21-1 Osawa, Mitaka, Tokyo 181-8588, Japan} 
\affil{${}^{3}$\ Department of Astronomy, Graduate School of Science, The University of Tokyo, 7-3-1 Hongo, Bunkyo-ku, Tokyo 113-0033, Japan}
\affil{${}^{4}$\ Astrobiology Center of NINS, 2-21-1, Osawa, Mitaka, Tokyo 181-8588, Japan}
\affil{${}^{5}$\ Korea Astronomy and Space Science Institute,  Daedeokdaero 776, Yuseong, Daejeon 305-348, South Korea} 
\affil{${}^{6}$\ Department of Astronomical Science, SOKENDAI (The Graduate University for Advanced Studies), 2-21-1 Osawa, Mitaka, Tokyo 181-8588, Japan}


\begin{abstract}
We have investigated the TUKH122 prestellar core in the Orion A cloud using ALMA 3 mm dust continuum, N$_2$H$^+$ ($J=1-0$), and CH$_3$OH ($J_K=2_K-1_K$) molecular line observations.
Previous studies showed that TUKH122 is likely on the verge of star formation because the turbulence is almost dissipated and chemically evolved among other starless cores in the Orion A cloud.
By combining ALMA 12-m and ACA data, we recover extended emission with a resolution of $\sim5\arcsec$ corresponding to 0.01 pc and identify 6 condensations with a mass range of $0.1-0.4$ $M_\odot$ and a radius of $\lesssim0.01$ pc.
These condensations are  gravitationally bound following a virial analysis and are embedded in the filament including the elongated core with a mass of $\sim29$ $M_\odot$ and a radial density profile of $r^{-1.6}$ derived by {\it Herschel}.
The separation of these condensations is $\sim0.035$ pc, consistent with the thermal jeans length at a density of $4.4\times10^5$ cm$^{-3}$. This density is similar to the central part of the core. 
We also find a tendency that  the N$_2$H$^+$ molecule seems to deplete at the dust peak condensation.  This condensation may be beginning to collapse because the linewidth becomes broader. 
Therefore, the fragmentation still occurs in the prestellar core by thermal Jeans instability and multiple stars are formed within the TUKH122 prestellar core.
The CH$_3$OH emission shows a large shell-like distribution and surrounds these condensations, suggesting that the CH$_3$OH molecule formed on dust grains is released into gas phase by non-thermal desorption such as photoevaporation caused by cosmic-ray induced UV radiation.

\end{abstract}





\section{Introduction}

To understand star formation processes, it is of great importance to reveal the initial conditions of star formation.
In nearby dark clouds at a distance of $\sim100$ pc, many observations have been performed using molecular lines and continuum emission and identified dense cores (or molecular dense cores) as birth places of stars \citep[e.g.,][]{mye83,ben89,war94,oni02,cas02}.
One of the most important information to understand the dense core properties is the density structure.
On the basis of near-infrared observations, \citet{alv01} showed that the Bonnor-Ebert Sphere model \citep{1956MNRAS.116..351B,1955ZA.....37..217E} can explain the observed radial column density profile of the Barnard 68.
Many nearby dense cores have also shown the Bonnor-Ebert structure \citep{2005AJ....130.2166K}.
The Bonnor-Ebert shape consists of a flat central region surrounded by a steeper outer region of $\rho\propto r^{-2}$.  
 Such a simple dense core is suggested to form a single protostar, binary stars, and multiple stars  by gravitational collapse \citep[e.g.,][]{mat03}.

The dense cores may be the most simplified description to determine the star-forming process and is often assumed as the initial conditions for simulations.
Therefore, the simple density structure can be applied to these isolated objects except for multiple star systems \citep[e.g.,][]{tok14,tok16,dun16,kir17}.
Recently, \citet{pin10,pin11,pin15} have discovered the quiescent thermal dense core, B5 in Perseus, showing a quadruple star system inside the core with the NH$_3$ high resolution observations. 
They suggested that the multiple star system is formed because the coherent core is fragmented into dense filaments with length of 5000 au.
Taking into account these results, the fragmentation in prestellar stage of dense cores may be needed to be a multiple star system.
Therefore, it is important to reveal the smallest and densest parts of the dense cores where stars are born.

However, the properties of dense cores in various star-forming environments have not been fully explored yet.
The majority of stars are thought to be produced through the formation of clusters \citep{lad03}.
In particular, ``giant molecular clouds (GMCs)'' are well known to be major sites of star formation in our Galaxy, and often show star cluster formation including massive stars.
Therefore, it is essential to observe the dense cores embedded in GMCs to reveal the initial conditions of star formation.
Furthermore, the observational targets should be dense cores on the verge of star formation.

To search for  dense cores with the initial conditions of star formation in GMCs,  chemical evolution may be one of the powerful tools to determine the evolutionary stages of the dense cores.
A pioneering study on the chemical evolution in dark clouds was done by \citet{1983ApJ...272..579S} and \citet{suz92}. 
They suggested that carbon chain molecules such as CCS, HC$_3$N, and HC$_5$N are abundant in dense cores in the early stages of star formation because
reactions involving atomic carbon or ionized carbon are effective in early stage and they are easily depleted onto dust in later stage \citep{aik01}. They also suggested
that the N-bearing molecule NH$_3$ is abundant in dense cores in the later stage
because N$_2$ molecule (precursor to N-bearing molecules) is slowly formed. 
 \citet{tat93,tat14a} showed that $N$(N$_2$H$^+$)/$N$(CCS) may indicate the chemical evolutionary stage.
Similarly,  \citet{oha14} found that the NH$_3$/CCS  column density ratio is anti-correlated with the CCS linewidth, and suggested that chemical evolution and turbulence dissipation can be indicators of the dynamical evolution of cores.
Therefore, if we can identify dense cores that are rich in N-bearing molecules and poor in carbon-chain molecules without
protostars, these cores will tell us the initial conditions of star formation.

\begin{figure}[htbp]
  \begin{center}
  \includegraphics[width=8.5cm,bb=0 0 842 595]{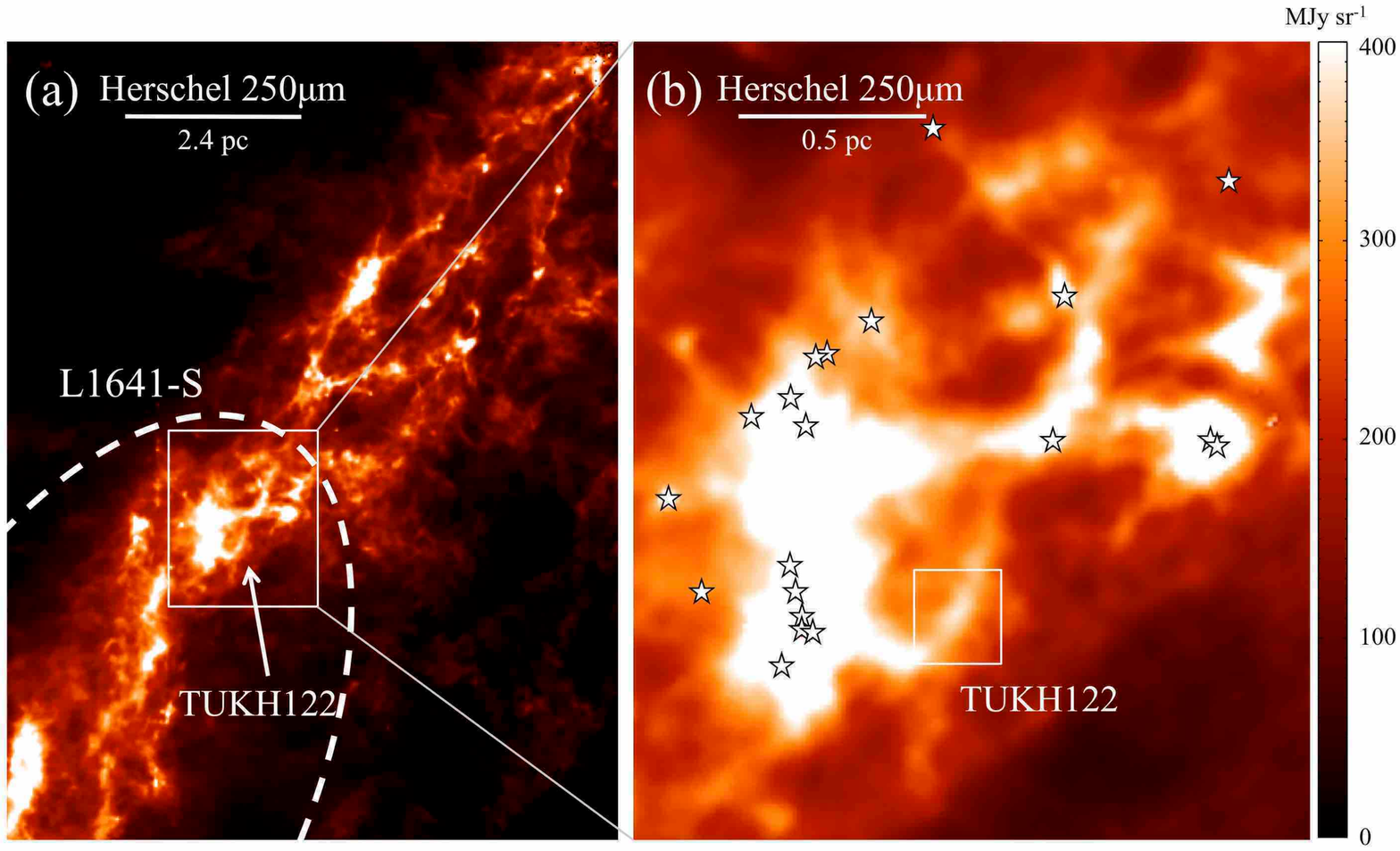}
  \end{center}
  \caption{The  {\it Herschel} 250 $\mu$m images. The left panel (a) shows a wide field and indicates L1641-S region. The right panel (b) shows the zoom up toward TUKH122. The open stars represent the locations of protostars identified by {\it Spitzer} \citep{meg12}.
  }
  \label{l1641}
\end{figure}

Based on these studies,  such cores are searched for in the Orion A cloud, the Vela C molecular cloud complex, and Planck cold clumps \citep{oha14,oha16p,tat14a,tat17}.
In particular, the Orion A cloud is a good target because it is one of the nearest GMCs from the earth and has been well studied.
\citet{tat14a} have found a dense core (TUKH122) having the largest value of $N$(N$_2$H$^+$)/$N$(CCS) among the starless dense cores in the Orion A cloud.
TUKH122 has $N$(N$_2$H$^+$)/$N$(CCS) $\sim3$ (typical starless dense cores have $\sim1.5$) and is located in the L1641 South region (see Figure \ref{l1641}). TUKH122 could be the dense core closest to star formation in our samples.
Note that no IRAC, MIPS, or SDSS sources are detected toward TUKH122, suggesting that it is in the prestellar phase.

Detailed observations have been performed toward TUKH122 with the VLA and  Nobeyama 45 m telescope.
The VLA NH$_3$ observations have identified an oval core with a size of 0.1 pc and condensations with a size of 0.03 pc embedded in the parent CS clump. The linewidth of NH$_3$ is narrow ($\sim0.2$ km s$^{-1}$), and both  the core and condensation are gravitationally bound \citep{tat14b}.
The single pointing NH$_3$ observations covering the whole main core with 0.05 km s$^{-1}$ velocity resolution identified not only the thermal narrow component but also a turbulent component, suggesting that the sharp transition from the parent clump to the quiescent dense core \citep{oha16}.
These results suggested that the TUKH122 core is on the verge of star formation and is one of the good targets to investigate the physical conditions and fragmentation process if this core forms multiple stars.
However, these observations were not enough to reveal the density distribution or fragmentation due to the resolution and sensitivity.

Recently, the {\it Herschel Space Observatory} has revealed that filaments are ubiquitous in star-forming clouds and dense cores are embedded in the filaments \citep[e.g.,][]{and10,arz11,hil11}. 
Therefore, TUKH122 is also important to understand dense core formation in the filaments in GMCs.

In this paper, we report ALMA observations toward TUKH122 at Band 3 with $\sim5\arcsec$ beam.   Note that we will follow the nomenclature of \citet{oha16c} and refer to cores as an entity of $0.01-0.1$ pc. We also refer to a dense condensation as an entity of $\lesssim0.01$ pc.
 The distance of Orion A cloud is derived to be 418 pc \citep{kim08}.

\begin{figure*}[htbp]
  \begin{center}
  \includegraphics[width=18.5cm,bb=0 0 2498 895]{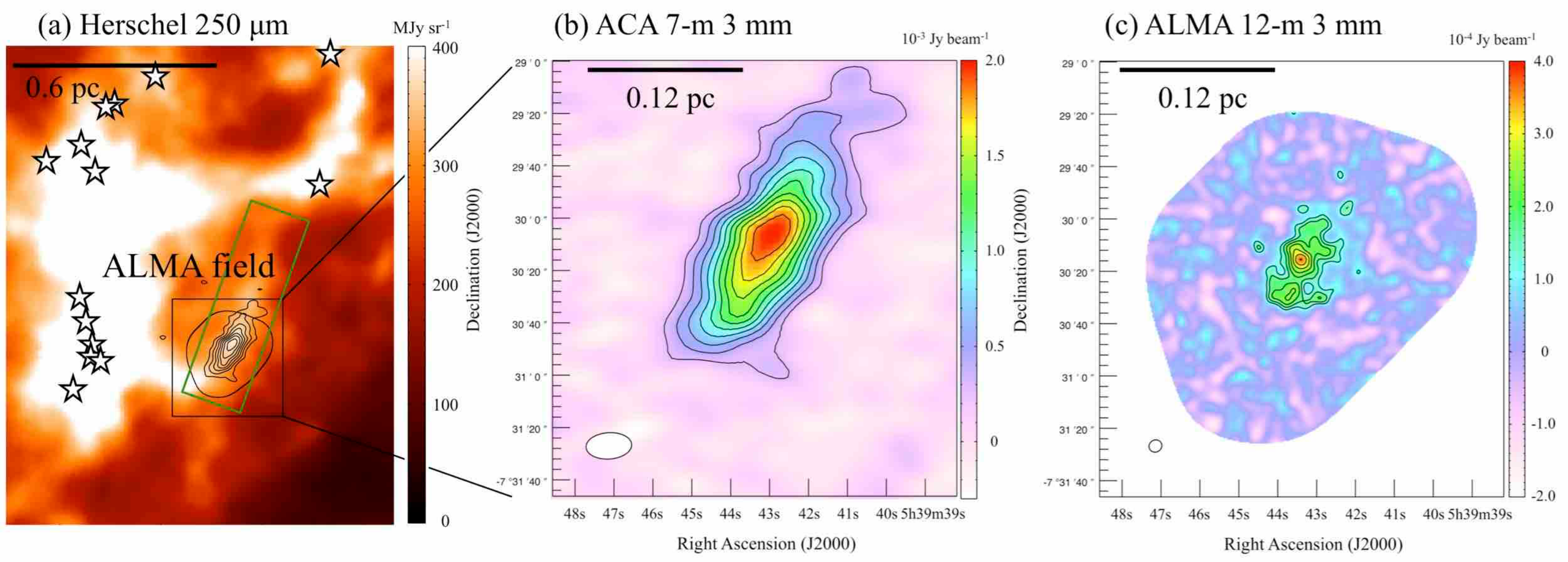}
  \end{center}
  \caption{(a) The {\it Hershcel} 250 $\mu$m dust continuum image toward the southern part of the Orion A cloud. The black contour shows the ALMA 12-m observed field toward TUKH122. The star symbols represent the protostars identified by {\it Spitzer}. The contours show the ACA 3 mm dust continuum emission as the same with panel (b). The green box shows the parent filamentary region.
  (b) The ACA 7-m 3 mm dust continuum image without the primary beam correction toward TUKH122 core.
The contours start at $3\sigma$ with intervals of $3\sigma$. The $1\sigma$ noise level is  87 $\mu$Jy beam$^{-1}$. The bottom left circle represents the beamsize of $17\arcsec\times10\arcsec$ (PA$=-86^\circ$) corresponding to $0.035$ pc $\times0.02$ pc. 
(c) The 3 mm dust continuum image without the primary beam correction toward TUKH122 core obtained by the ALMA 12-m array.
The contours start at $3\sigma$ with intervals of 1$\sigma$. The $1\sigma$ noise level is  43 $\mu$Jy beam$^{-1}$. The bottom left circle represents the beamsize of $5\farcs0\times4\farcs6$ (PA$=-61^\circ$) corresponding to $0.01$ pc $\times0.0094$ pc. The combined ALMA image is shown in Figure \ref{cont_combine}.}
  \label{map}
\end{figure*}

{In Section 2, we describe the observations and data reduction. In Section 3, we present and describe the dust continuum maps taken by the {\it Herschel}, ACA, and ALMA observations as well as the molecular line maps of the ACA and ALMA data. In Section 4, we discuss the formation processes of condensations in the core and possible scenario of multiple star systems. Our conclusions are summarized in Section 5.

\section{Observations}

\begin{deluxetable*}{p{70mm}lccc}
\tablewidth{0pt}
\tablecaption{Beam Sizes and Sensitivities}
\tablehead{
Parameters &	ALMA 12 m Array	& ACA 7 m Array & ALMA-ACA combined }
\startdata
Synthesized beam size of continuum observation & $\sim5\farcs0\times4\farcs6$	& $\sim17\arcsec\times10\arcsec$&	 $\sim5\farcs7\times5\farcs2$	\\
Synthesized beam size of line observation & $\sim3\farcs0\times2\farcs7$	& $\sim18\arcsec\times10\arcsec$&	 $\sim3\farcs2\times2\farcs9$	\\
Sensitivity of continuum observation (rms) &$\sim43$ $\mu$Jy beam$^{-1}$	&	$\sim87$ $\mu$Jy beam$^{-1}$	&$\sim40$ $\mu$Jy beam$^{-1}$\\
Sensitivity of line observation (rms)$^a$ &$\sim6.5$ mJy beam$^{-1}$	&	$\sim31$ mJy beam$^{-1}$	&$\sim6.5$ mJy beam$^{-1}$
\enddata
\tablenotetext{a}{The velocity resolution is $\sim0.098$ km s$^{-1}$.}
\end{deluxetable*}

TUKH122 was observed with the ALMA 12-m Array on 2016 March 10-12 in the C36-2/3 configuration with a total of 38 antennas and with the 7-m Array of the Atacama Compact Array \citep[ACA; also known as the Morita Array;][]{igu09} on 2016 May 28, August 27, and September 1-7 with a total of 10 antennas (Cycle 3 program, Project ID: 2015.1.01025.S).
The number of 12-m pointings for the mosaic mapping is five. The  Band 3 receivers were used.  The system temperature was in the range of 40 to 140 K.
The correlator was set to have four spectral windows (two windows for continuum, one for N$_2$H$^+$, and one for CH$_3$OH).
The velocity resolution of the molecular line observations was set to be $\sim0.098$ km s$^{-1}$.
The four quasars (J0423-0120, J0522-3627, J0854+2006, and J0510+1800) were observed for bandpass calibration. J0542-0913 was observed for phase calibration.
 Flux calibration was performed using J0423-0120, J0510+1800, J0522-3627, and Uranus.

The reduction and calibration of the data were done with CASA version 4.5.3 \citep{mcm07} in a standard manner. 
All images were reconstructed with the CASA task CLEAN using natural weighting. To improve the sensitivity, we also used uv tapering (50 k$\lambda$ for the molecular lines and 20 k$\lambda$ for continuum)  for CLEAN.
The pixel size was set to 0$\farcs$5. 
To make high spatial resolution images  recovering the extended emission, we combine the ALMA 12-m and ACA in the uv plane. 
The resultant spatial resolutions are written in each section.
The synthesized beams and the sensitivities of the dust continuum and the molecular lines are listed in Table 1.

\section{Results}

Figure \ref{map} shows the dust continuum maps toward the TUKH122 region at different scales with {\it Herschel}  and ALMA. 
The various observations with the different spatial resolutions reveal the different structures from filament to condensation. Figure \ref{map} (a) shows the location of TUKH122 on the {\it Herschel} 250 $\mu$m map \citep[Gould Belt Survey Archive:][]{and10,roy13,pol13} overlaid with ACA 3 mm dust continuum emission contours.
We find a filamentary structure along north-south direction with a length of $\sim0.6$ pc shown in the green box}, which is wider than ALMA observed field.

Figures \ref{map} (b) and (c) show the 3 mm dust continuum images obtained by ACA 7-m array and ALMA 12-m, respectively. The beamsize is $17\arcsec\times10\arcsec$ (PA$=-86^\circ$) for ACA and $5\farcs0\times4\farcs6$ (PA$=-61^\circ$) for ALMA 12-m. These images are not corrected by the primary beam pattern to keep the same noise level within the images.
The ACA observations show the simple elongated structure ($\sim0.12$ pc $\times$ 0.048 pc in {\cal FWHM}) toward TUKH122 region. 
In the 12-m array data (Figure \ref{map} c), compact condensations are clearly seen, whereas such components are not spatially resolved in the 7-m array data.  The ALMA-12m observations only reveal the small condensations within the core.

\begin{figure}[htbp]
  \begin{center}
  \includegraphics[width=8.5cm,bb=0 0 1649 1638]{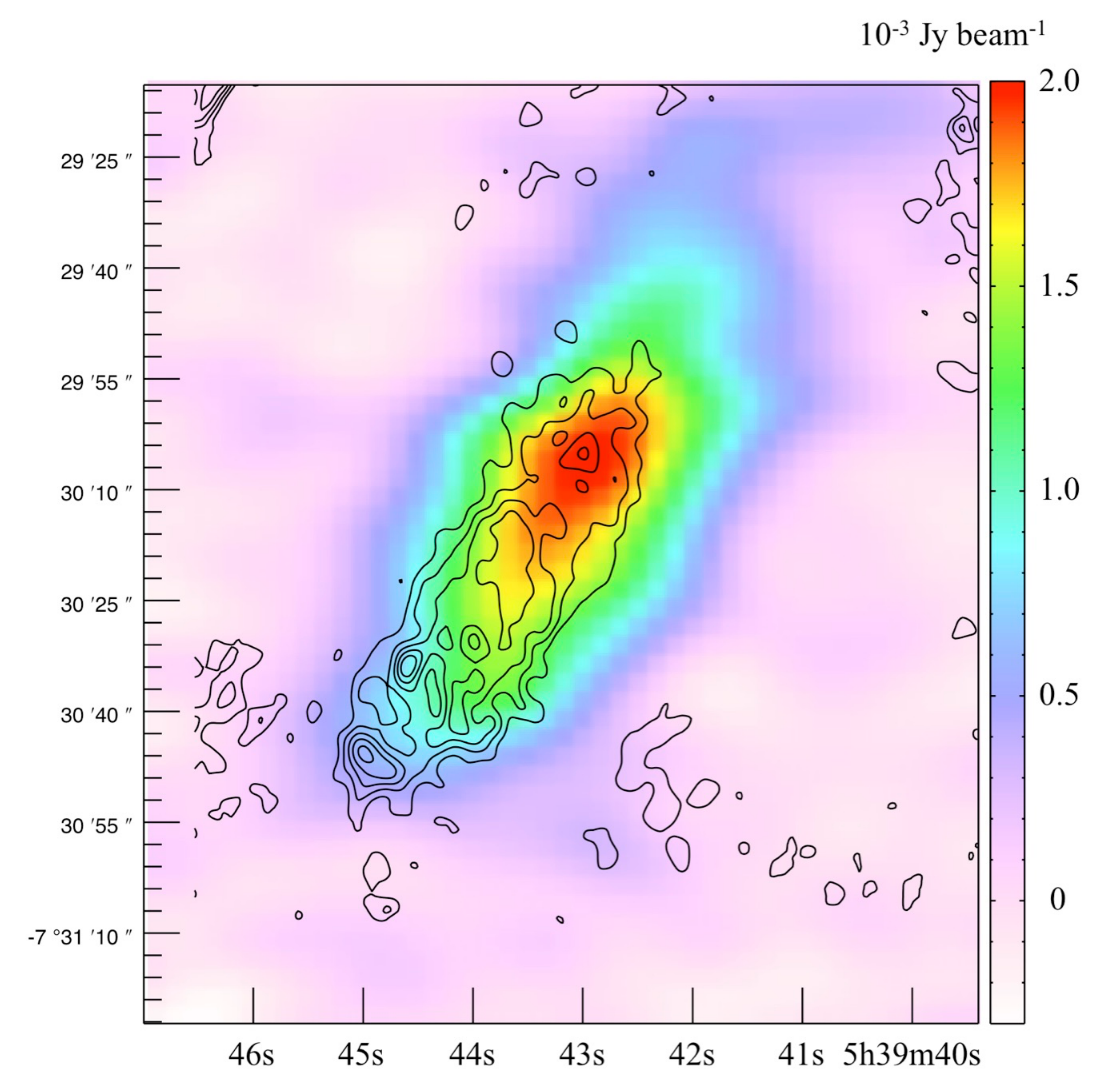}
  \end{center}
  \caption{ The color map shows ACA 7-m 3 mm dust continuum image without the primary beam correction toward TUKH122 core. The contours show NH$_3$ integrated intensity map. The lowest contour level and contour interval are 2.4 mJy beam$^{-1}$ km s$^{-1}$. }
  \label{vla_nh3}
\end{figure}

The oval structure of the TUKH122 core was already reported by \citet{tat14b} with VLA NH$_3$ observations. 
Figure \ref{vla_nh3} shows the ACA 7-m 3mm dust continuum image with the VLA NH$_3$ integrated intensity contours.
The 3 mm dust continuum emission taken by ACA is extended compared with the VLA NH$_3$ emission, which may be explained by filtering out in the VLA interferometric data.
These oval structures are consistent with the typical dense core found in the low-mass star forming region.

We suggest that the TUKH122 core is embedded in a parent filament having larger scale emission. However, it is difficult to distinguish the core emission component from the filament. 
Therefore, the {\it Herschel} and ACA observations trace the combination of the filament and the core components.
Most emission may be from the core because the TUKH122  core has been already formed.

\begin{figure}[htbp]
  \begin{center}
  \includegraphics[width=8.5cm,bb=0 0 1988 1873]{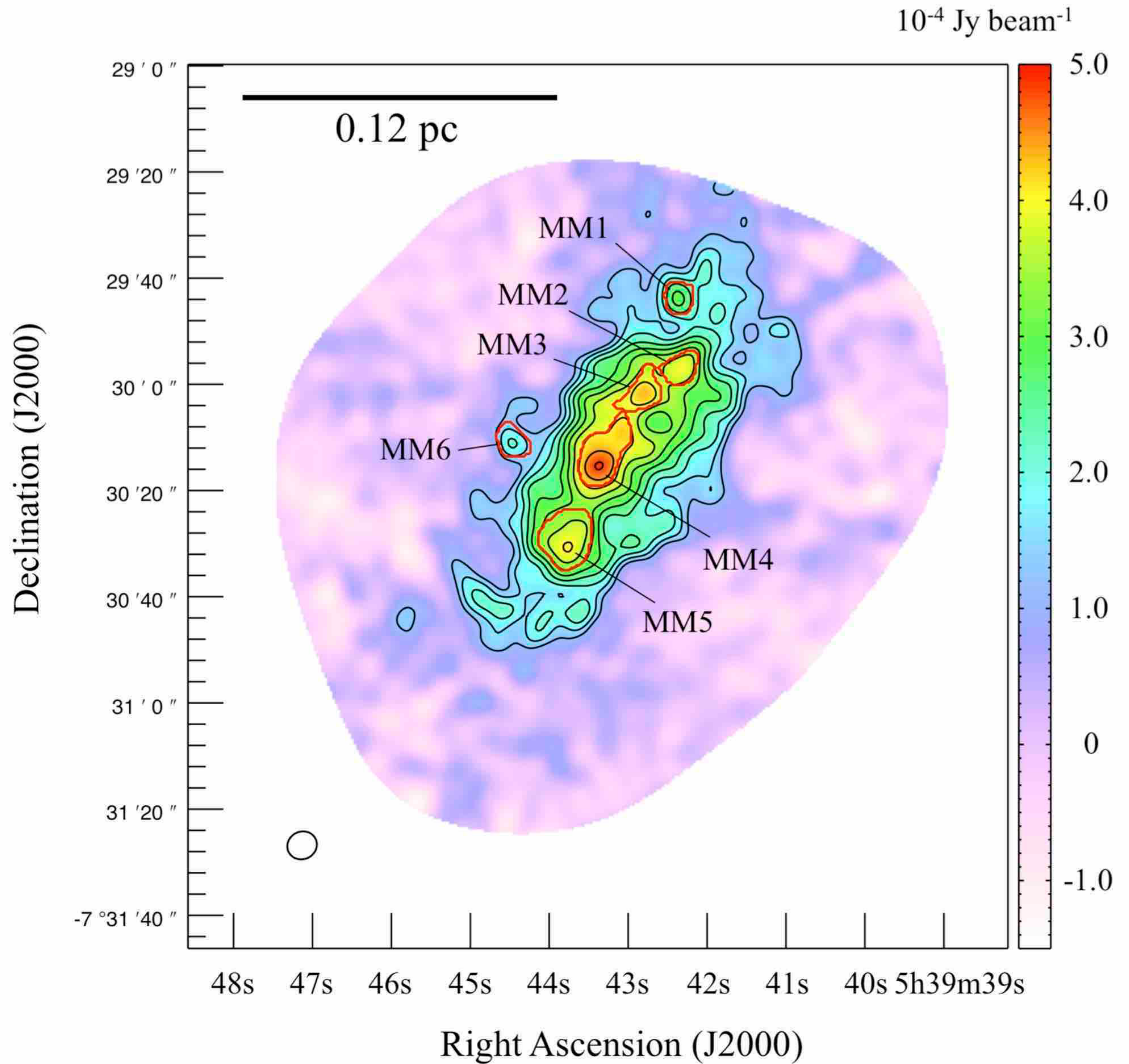}
  \end{center}
  \caption{The ALMA-ACA combined 3 mm dust continuum image toward TUKH122 core without the primary beam correction.
The contours start at $3\sigma$ with intervals of 1$\sigma$. The $1\sigma$ noise level is  40 $\mu$Jy beam$^{-1}$.
The red contours represent the condensations. The bottom left circle represents the beamsize of $5\farcs7\times5\farcs2$ (PA$=-66^\circ$).}
  \label{cont_combine}
\end{figure}

Figure \ref{cont_combine} shows the 3 mm dust continuum image obtained by the ALMA-ACA combined data.
The contours start at 3$\sigma$ with intervals of 1$\sigma$.  The 1$\sigma$ noise level is  40 $\mu$Jy beam$^{-1}$ and the beamsize is $5\farcs7\times5\farcs2$ (PA$=-66^\circ$).

\subsection{Analysis of the Herschel and ALMA continuum data}
To derive the dust temperature and mass of the ALMA-ACA observing region by using the {\it Hershcel} 250, 350, and 500 $\mu$m data, we adopt a dust opacity of 5.5 cm$^2$ g$^{-1}$ at 350 $\mu$m and 2.7 cm$^2$ g$^{-1}$ at 500 $\mu$m modeled by \citet{orm07}.
These values have been shown to reporduce well the observed SED from 2.2  to 850 $\mu$m in the Orion A cloud as suggested by \citet{lom14}.
The mass $M$, and dust temperature $T_d$, are computed with
\begin{equation}
M=\frac{F_\nu d^2}{\kappa B_\nu(T_d)}f_d,
\label{1}
\end{equation}
where $F_\nu$ is the observed flux in Jy, $d$ is the distance to the target, $B_\nu(T_d)$ is the Planck function, $\kappa$ is the dust opacity, and $f_d$ is the gas to dust mass ratio (assumed to be 100).
By calculating the flux density within the ALMA observed field, the dust temperature of TUKH122 is derived to be 12 K, consistent with the previous results of $T_{\rm kin}=11$ K obtained by NH$_3$ spectra \citep{oha16}.  The mass is also derived to be $\sim29$ $M_{\odot}$.
In spite of the observational studies that dust temperature is often higher than gas temperatures because of averaging the line-of-sight of the dust, the similar temperatures between gas and dust may suggest that this region is widely isothermal.

\begin{figure}[htbp]
  \begin{center}
  \includegraphics[width=8.5cm,bb=0 0 1881 1726]{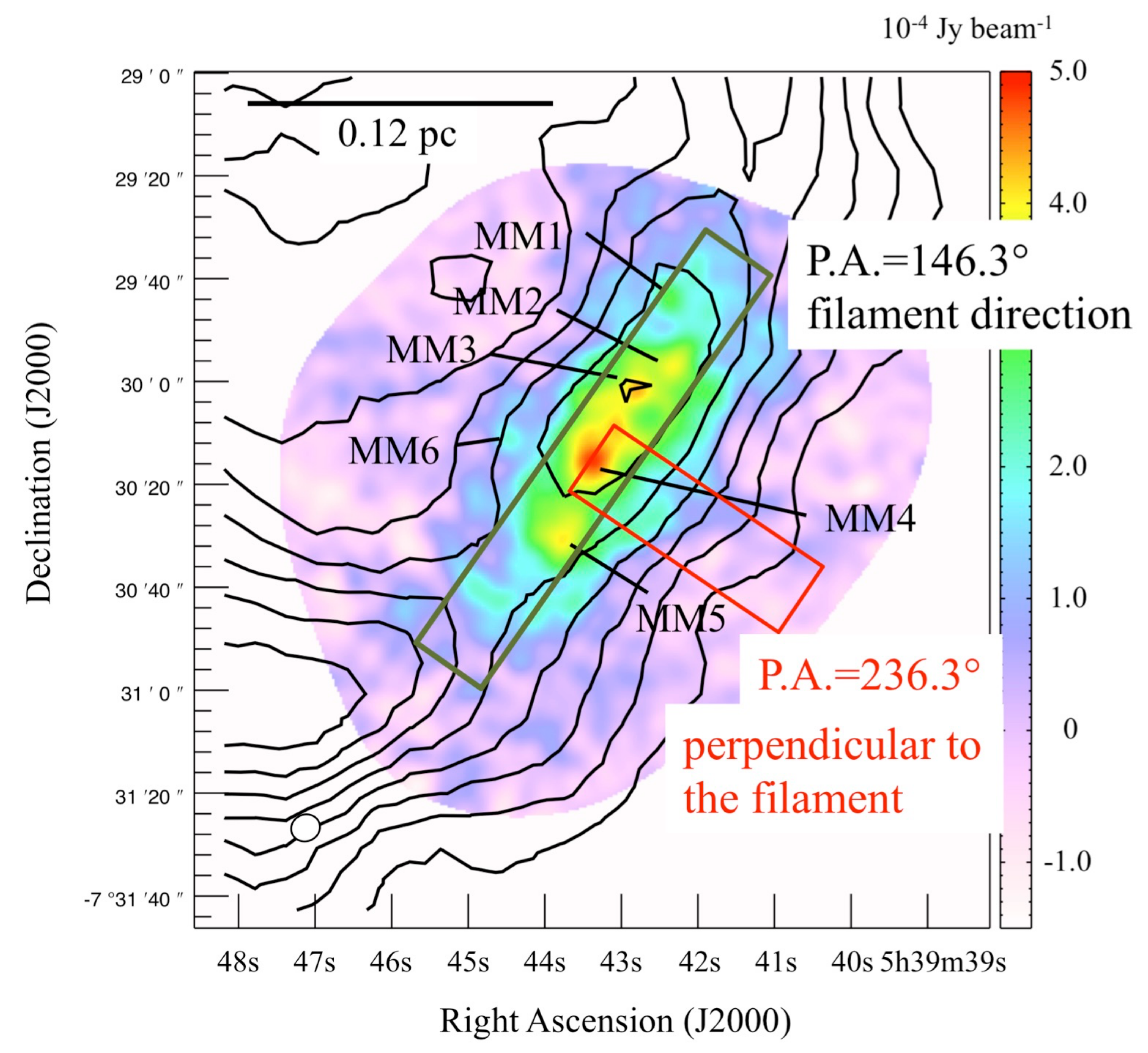}
  \end{center}
  \caption{The color map shows the same as Figure \ref{cont_combine}. The contours show the {\it Herschel} 250 $\mu$m dust continuum and start at 260 MJy sr$^{-1}$ with 20 MJy sr$^{-1}$ intervals.  The directions of the parallel and perpendicular to the filamentary structure are also shown with a width of 15$\arcsec$.
  }
  \label{cont_combine_pa}
\end{figure}

The plane of the sky orientation of the filament and elongated core is measured to be a position angle of $\rm P.A. =146.3$ degrees derived by 2D gaussian fitting on the {\it Herschel} image.
The directions of the parallel and perpendicular to the filamentary structure are shown in Figure \ref{cont_combine_pa}.

\begin{figure}[htbp]
  \begin{center}
  \includegraphics[width=8.5cm,bb=0 0 2501 1804]{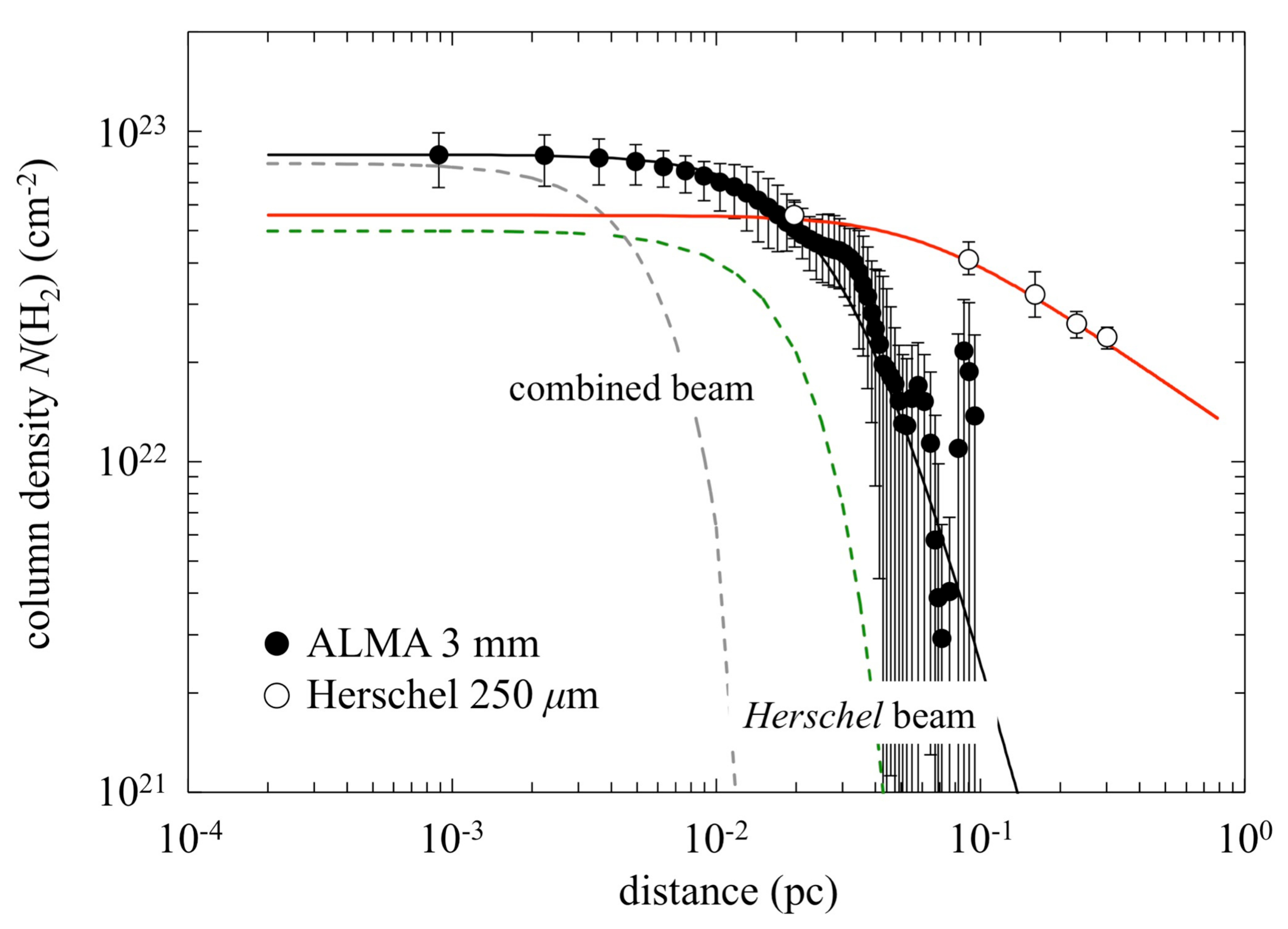}
  \end{center}
  \caption{The column density profiles perpendicular to the filament direction taken by ALMA-ACA observations and the {\it Herschel} 250 $\mu$m observations. Error bars show the maximum and minimum distributions of the profiles in each region. The black and red lines represent the Plummer like function with the power laws of 4.0 and 1.6, respectively. The dash lines represents the beam patterns of the ALMA-ACA combined data and {\it Herschel} 250 $\mu$m data.}
  \label{perpen}
\end{figure}

To investigate the density structure of this region, Figure \ref{perpen} shows the column density profiles perpendicular to the filamentary structure.
The column density profiles are derived with the ALMA-ACA and {\it Herschel} observations by using the equation (\ref{1}) assuming the surface are of the  ALMA-ACA beamsize of $\sim5\farcs2$ and the {\it Herschel} beamsize of $\sim18\arcsec$.
We measure a radial column density profile to the south-westen direction ($\rm P.A. =236.3^\circ$) perpendicular to the filament.
Then, the profile is calculated by averaging the flux densities within 15$\arcsec$ width at each distance from the peak position (see also Figure \ref{cont_combine_pa}).
We do not use the data of the north-eastern side because several protostars may affect the density and temperature in this side (see Figure \ref{map}).

Figure \ref{perpen} shows that the {\it Herschel} observations recover extended emission at a distance larger than 0.1 pc, while ALMA-ACA observations only detect dense gas within 0.1 pc.
This is consistent with the results that the ALMA 12-m and ACA 7-m observations are not sensitive to the extended envelope for scale larger than 0.1 pc.
We also find that the column densities of the ALMA-ACA observations are higher than that of the {\it Herschel} observations inside 0.02 pc.
This would most likely be caused by the low resolution of the {\it Herschel} observation because the ALMA-ACA observations are three times better spatial resolution than {\it Herschel}. For example,  {\it Herschel} does not resolve the 0.04 pc scale condensation found in this study.

To understand the density structure of the parent filament with the core, we plot the Plummer like function as the red and black lines in the figure.
The Plummer-like profile is defined as
\begin{equation}
\rho(r)=\frac{\rho_c}{[1+(r/R_f)^2]^{p/2}},
\label{plummer}
\end{equation}
and the corresponding surface density profile is
\begin{equation}
\Sigma(r)=A\frac{\rho_cR_f}{[1+(r/R_f)^2]^{(p-1)/2}},
\label{plummer2}
\end{equation}
where $A$ is a numerical factor. 
The numerical factor of $A$ depends on the density slope $p$. $A=\pi$ corresponds to be $p=2$, while $A=\pi/2$ corresponds to be $p=4$ \citep{arz11}.
We derive the column density assuming the mean molecular weight of 2.33.
The profile is determined by the central density $\rho_c$, the radius $R_f$, and the density slope $p$ \citep[][]{nut08,arz11}.
Note that the isothermal hydrostatic filament has $p=4$ \citep{sto63,ost64}.
Recently the {\it Herschel} observations suggest $p\sim2$ \citep[e.g.,][]{arz11} and that the dynamical contraction makes a flatter density slope. 
In Figure \ref{perpen}, the red line shows the power law of $p=1.6$, while the black solid line shows the power law of $p=4.0$, nicely fitting the data plotted.
The FWHM of the {\it Herschel} data fitting is  0.16 pc.
On the other hand, the ALMA-ACA observations show a thinner and steeper filament due to the missing flux.

The slope of $p=4.0$  is consistent with the isothermal hydrostatic filament or the density profile of prestellar cores \citep{whi01}. 
However, the ALMA-ACA observations only recover 8.4 $M_\odot$ out of a total of $\sim29$ $M_\odot$.
It is difficult to distinguish the density structure of the TUKH122 core from the parent filament by these observations due to the missing flux.
We cannot discuss the density profile or the power law of   the ALMA-ACA observations.
We only show that the missing flux is not the background emission but the surrounding gas associated to the filament.

It is possible that the power law of $p=1.6$ indicates the dynamical contraction of the filament rather than the background effect.
The filament actually forms the dense core and condensations.
Therefore, we indicate that the power law of $p=1.6$ is the radial profile of the parent filament that already forms the elongated core.

\subsection{The 3 mm dust continuum emission seen by ALMA}

To identify dense condensations precisely, we apply the dendrogram method \citep{ros08} to the combined 3 mm dust continuum image.
The dendrogram traces local maxima and describes hierarchical structures, which have been  shown  to  be  more  robust  against  noise  and  user defined parameters \citep[e.g.,][]{goo09,pin09}. 
This method gives three types of structures: leaf, branch, and trunk. We define ``leaves'' as ``condensations'' in this paper because the leaf is the smallest structure.
We adopt a threshold of 3$\sigma$ and 1$\sigma$ interval steps.
The minimum number of pixels are $3\times\theta_{\rm beam}$.  
In addition, we select the condensations as the robust one when they can be identified as detection above  3$\sigma$ with ALMA 12-m data.
The identified condensations are shown in Figure \ref{cont_combine} as the red contours and listed in Table \ref{dust}.

\begin{deluxetable*}{p{8mm}ccccccccccc}
\tablewidth{0pt}
\tablecaption{Physical Parameters of dense condensations \label{dust}}
\tablehead{
Sources &	R.A.	& Decl& $S_{\rm peak}^{\rm a}$ & $S_{\rm int}^{\rm a}$& Mass$^{\rm b}$ &  $r^{\rm c}$	& $n$ &	Velocity$^{\rm d}$ & linewidth$^{\rm d}$ & Virial Mass \\
& (h:m:s)	&(d:m:s) 	&(mJy beam$^{-1}$)	&(Jy)& ($M_\odot$) &($\arcsec$)&  (cm$^{-3}$) &(km s$^{-1}$)	& (km s$^{-1}$)&($M_\odot$)  
}
\startdata
MM-1 &5:39:42.34	&	-7:29:43	&	0.45	&	0.33	&		0.12	&	1.6	&$1.5\times10^7$ &3.81	&  0.20 &0.03	\\
MM-2 &5:39:42.30&	-7:29:56	&	0.46	&	0.43	&		0.15	&	2.0	& $9.2\times10^6$ &	3.76& 0.22  &0.04 \\
MM-3 &5:39:42.78	&	-7:30:01	&	0.46	&	0.55	&		0.20	&	2.7	&$5.2\times10^6$  &3.80	&0.20  &0.05\\
MM-4 &5:39:43.37	&	-7:30:15	&	0.48	&	1.1	&		0.39	&	4.5	&$2.2\times10^6$ &	3.77& 0.38  &0.3\\
MM-5 &5:39:43.76&	-7:30:31	&	0.42	&	0.99	&		0.36	&	4.7	&$1.8\times10^6$  &	3.79&  0.22&0.1\\
MM-6 &5:39:44.48	&	-7:30:10	&	0.22	&	0.17	&		0.06	&	1.9	& $4.9\times10^6$ &3.66	&0.23   &0.04
\enddata
\tablenotetext{a}{Fluxes are corrected for the primary beam attenuation.}
\tablenotetext{b}{The mass is derived from the combined map.}
\tablenotetext{c}{The deconvolved size of the radius.}
\tablenotetext{d}{The velocity and linewidth are derived by Gaussian fitting of CH$_3$OH.}
\end{deluxetable*}

\begin{figure*}[htbp]
  \begin{center}
  \includegraphics[width=16cm,bb=0 0 2794 1316]{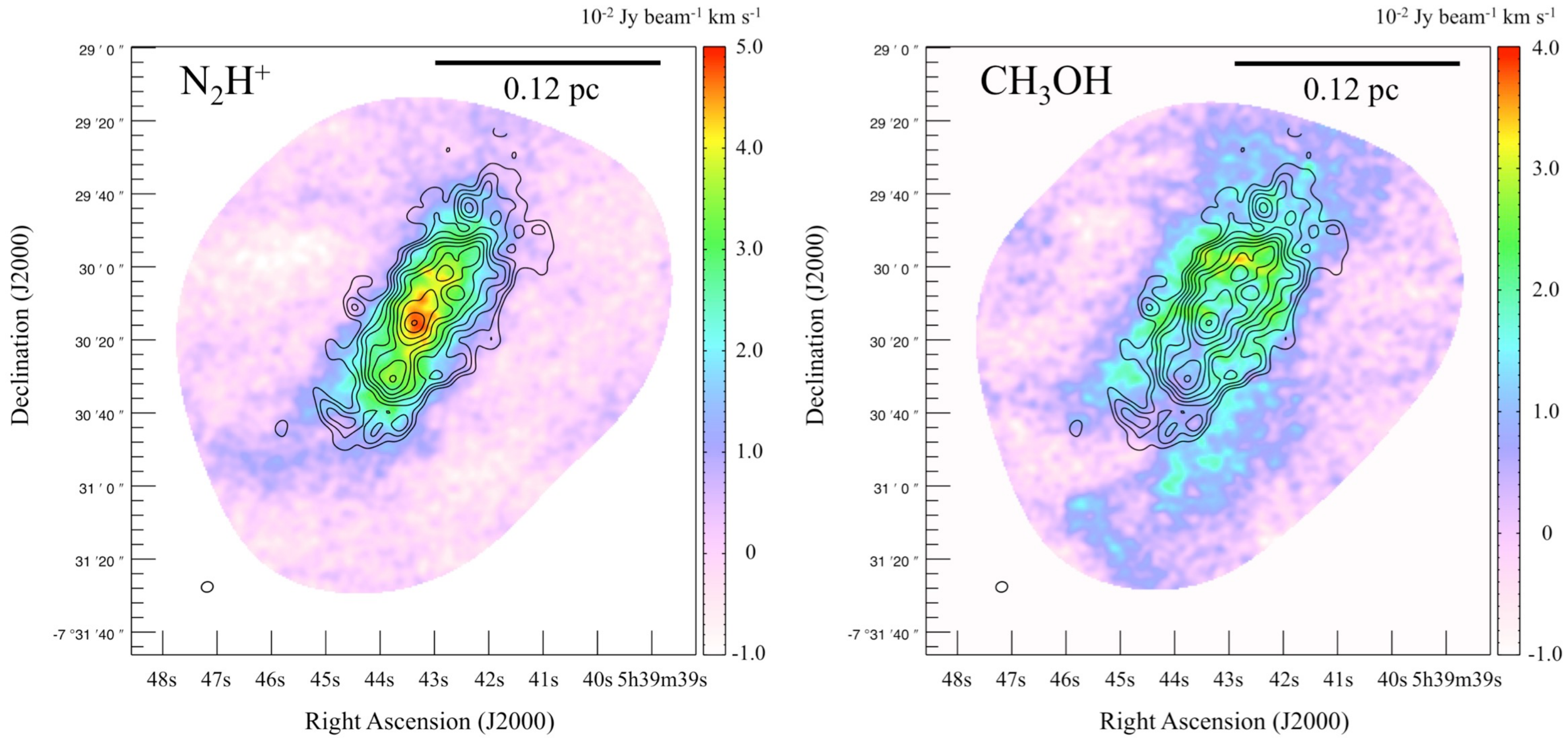}
  \end{center}
  \caption{(Right): The N$_2$H$^+$ velocity-integrated intensity image toward TUKH122 core obtained form ALMA-ACA combined data.
  The velocity range is $3.4-4.0$ km s$^{-1}$. The contours represent the 3 mm dust continuum map and are the same as Figure \ref{cont_combine}. The bottom-left circle represents the beamsize of 3$\farcs2\times2\farcs9$ (PA$=-73^\circ$)  (Left): The CH$_3$OH velocity-integrated intensity image toward TUKH122 core obtained form ALMA-ACA combined data.
  The velocity range is $3.4-4.0$ km s$^{-1}$. The contours represent the 3 mm dust continuum map and are the same as Figure \ref{cont_combine}.
  The bottom-left circle represents the beamsize of 3$\farcs2\times2\farcs8$ (PA$=-72^\circ$) .}
  \label{integ}
\end{figure*}

The mass of the 3 mm dust condensations are computed with the equation (\ref{1}) at a dust temperature $T_d=12$ K.
We adopt a dust opacity of 0.0755 cm$^2$ g$^{-1}$, assuming a dust emissivity index of $\beta=2$ \citep{orm07}.
As a result, the condensation mass ranges from 0.1 to 0.4 $M_\odot$.  The typical deconvolved radius is $r\sim2\farcs9$. The radius is calculated from the surface area of the identified condensations. The H$_2$ densities are $n\sim10^{6-7}$ cm$^{-3}$ which are of the order of or higher than those of typical low-mass prestellar cores \citep[e.g.,][]{oni02,eno08}.
The condensation in the center of TUKH122, MM4, is the most massive and has $\sim0.4$ $M_\odot$.
Note that the deconvolved radius of $r\sim2\farcs9$ is very close to the beamsize radius. The condensations MM1,2 and 6 show that the peak intensities have larger values than the total fluxes (see also Table \ref{dust}).
These results indicate that these three condensations (MM1,2 and 6) are not well resolved with our observations.
However, the distances between the condensations are larger than the beamsize, indicating that we can determine the locations of the condensations with our observations. 
We find an interesting trend that the northern condensations (MM1, MM2, and MM3) have $0.1-0.2$ $M_\odot$, while the southern condensations (MM4 and MM5) have $\sim0.4$ $M_\odot$.
We discuss the different mass distribution along the filament in the following section.

The total mass of TUKH122 is derived to be 8.4 $M_\odot$ (2.5 $M_\odot$ for ALMA 12-m data) using the ALMA-ACA combined data. 
Taking into account the total mass of TUKH122 ($\sim29$ $M_\odot$) estimated by {\it Herschel}, there is still a large fraction of flux that is not recovered by ALMA.
In order to check how the condensation masses vary, we combine our ALMA and ACA data with the {\it Herschel} 250 $\mu$m data using {\it feather} in CASA. 
Note that we rescale the 250 $\mu$m flux to the 3 mm flux assuming the dust temperature of 12 K and ice-covered silicate-graphite conglomerate grains with a dust emissivity index of $\beta=2$  modeled by \citet{orm07}. We confirm that the total mass is recovered to be $\sim29$ $M_\odot$.
Then, we apply the dendrogram method to the 3 mm continuum image in the same way as the above. 
We find that the condensation mass ranges from 0.1 to 0.7 $M_\odot$. 
The MM1 to MM4 condensations increase in mass only $\sim10$\% but the MM5 increases $\sim60$\% in mass, suggesting that the condensation masses do not change even when we add the total power except for MM5.
However, we should also note that if condensations have relatively extended structures as MM5 does, the masses may be underestimated due to the missing flux.

We also study the dynamical stability of these condensations.
Assuming a uniform density structure, the virial mass can be estimated as
\begin{equation}
M_{\rm vir}=210 \times \left( \frac{r}{ \mathrm{pc}} \right) \times \left( \frac{\Delta v}{ \mathrm{km \, s^{-1}}} \right)^2 M_\odot,
\label{4}
\end{equation}
\\
where $r$ is the radius and $\Delta v$ is the linewidth \citep{mac88}.
The linewidth is derived by gaussian fitting of the CH$_3$OH ($J_K=2_0-1_0$ $A^{+}$) emission because this line is optically thin.
The centroid velocity, linewidth, and virial mass are shown in Table \ref{dust}.
As shown in the table, the condensation mass is much higher than the virial mass.
Therefore, these condensations are not in virial equilibrium and may collapse immediately unless the magnetic field counterbalances gravity.

\subsection{The molecular line emission of N$_2$H$^+$ and CH$_3$OH}

Figure  \ref{integ} shows in colors the velocity-integrated intensity maps of N$_2$H$^+$ ($J=1-0$, $F_1=1-1$, $F=0-1$) and CH$_3$OH ($J_K=2_0-1_0$ $A^{+}$) emission without the primary beam corrections. The ALMA 12-m and ACA 7-m data are combined.
The $F_1=1-1$, $F=0-1$ transition line is the weakest hyperfine component, and hence, it's optical depth is the smallest.
The contours are the 3 mm dust continuum (same as Figure \ref{cont_combine}). 
The 3 mm dust continuum and the N$_2$H$^+$  transition maps show that the intensities concentrate at the central part and have the same peak position.
On the other hand, the CH$_3$OH emission is weak at the central part of the TUKH122 core, resembling a shell-like structure that surrounds the continuum and N$_2$H$^+$ emission.
These configurations are quite similar to the starless cores of L1498 and L1517B, and L1544 \citep{taf06,vas14,spe16}, indicating that
the CH$_3$OH and N$_2$H$^+$ molecules have chemical differentiation.

The CH$_3$OH molecule is suggested to be formed on the surface of dust grains via successive hydrogenation of CO \citep[e.g.,][]{wat02} and is released by external heating and/or sputtering due to star formation activity.
Therefore, the CH$_3$OH lines are often used as a tracer of star formation activity  and shocks.
However, even in the cold quiescent clouds including this object, the CH$_3$OH lines have also been observed with moderate intensities. \citet{vas14} suggested that the non-thermal desorption mechanism is important for the observed emission of CH$_3$OH. \citet{som15}  also suggested that the CH$_3$OH formed on dust grains is liberated into the gas phase by non-thermal desorption such as photoevaporation caused by cosmic-ray induced UV radiation.
According to this  mechanism, the column density of gaseous CH$_3$OH is proportional to the path length of emitting region along the line of sight. 
Because the UV radiation is shielded in the core interiors, the emitting region has a shell-like structure.
The extended CH$_3$OH distribution seems to be consistent with this scenario.
Note that the CH$_3$OH shell structure is mainly observed in the north region and absent in the southeast. This may be because the CH$_3$OH emission is filtered out in the interferometric data if the CH$_3$OH emission is widely distributed or the CO molecule (the precursor of the CH$_3$OH) is less abundant on dust grains in the south region if the southern part is chemically young (CO has not been formed yet).

\citet{tat14b} also observed TUKH122 core with the NH$_3$ and CCS molecules using VLA and found that the CCS emission surrounds the NH$_3$ core.
Figure \ref{ccs-ch3oh} shows the velocity-integrated intensity map of the ALMA CH$_3$OH observations overlaid with the contours of the VLA CCS observations.
In comparison with these distributions, we find that the CCS emission is located on the CH$_3$OH shell structure or slightly outside the shell.

\begin{figure}[!htbp]
  \begin{center}
  \includegraphics[width=8.5cm,bb=0 0 1759 1690]{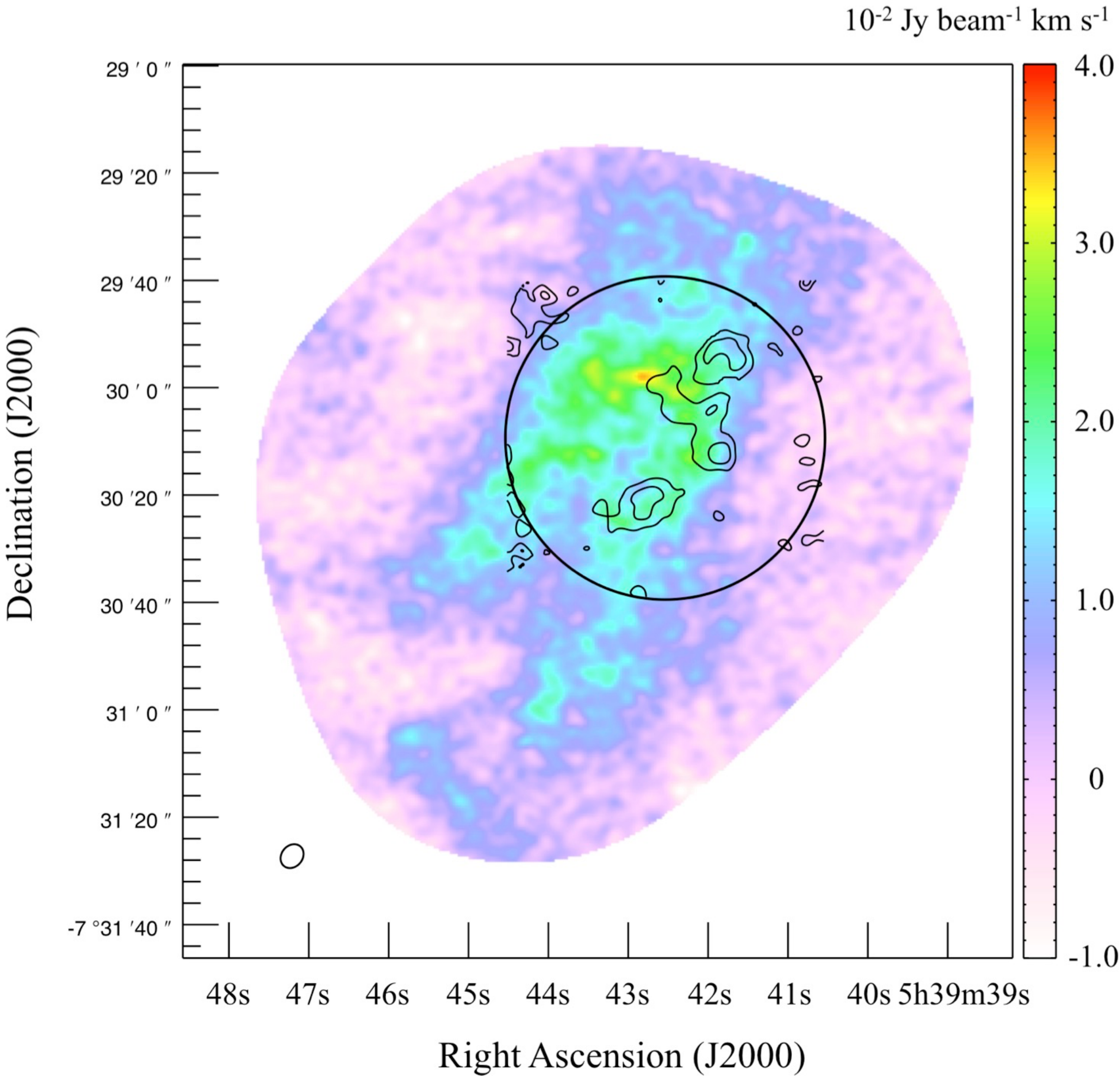}
  \end{center}
  \caption{The ALMA CH$_3$OH integrated intensity map (color) superimposed on the VLA CCS integrated intensity map (contours). The large circle delineates the primary beam size of the VLA interferometric observations. The contour interval is 6.9 mJy beam$^{-1}$ km s$^{-1}$, which corresponds to 3$\sigma$ and 1.5$\sigma$ at the map center and at the edge of the primary beam, respectively. In the lower left corner, the synthesized beam of 4$\farcs7\times3\farcs9$ (PA$=-40^\circ$) for CCS is shown.}
  \label{ccs-ch3oh}
\end{figure}

\begin{figure*}[!t]
  \begin{center}
  \setcounter{figure}{9}
  \includegraphics[width=19cm,bb=0 0 2938 975]{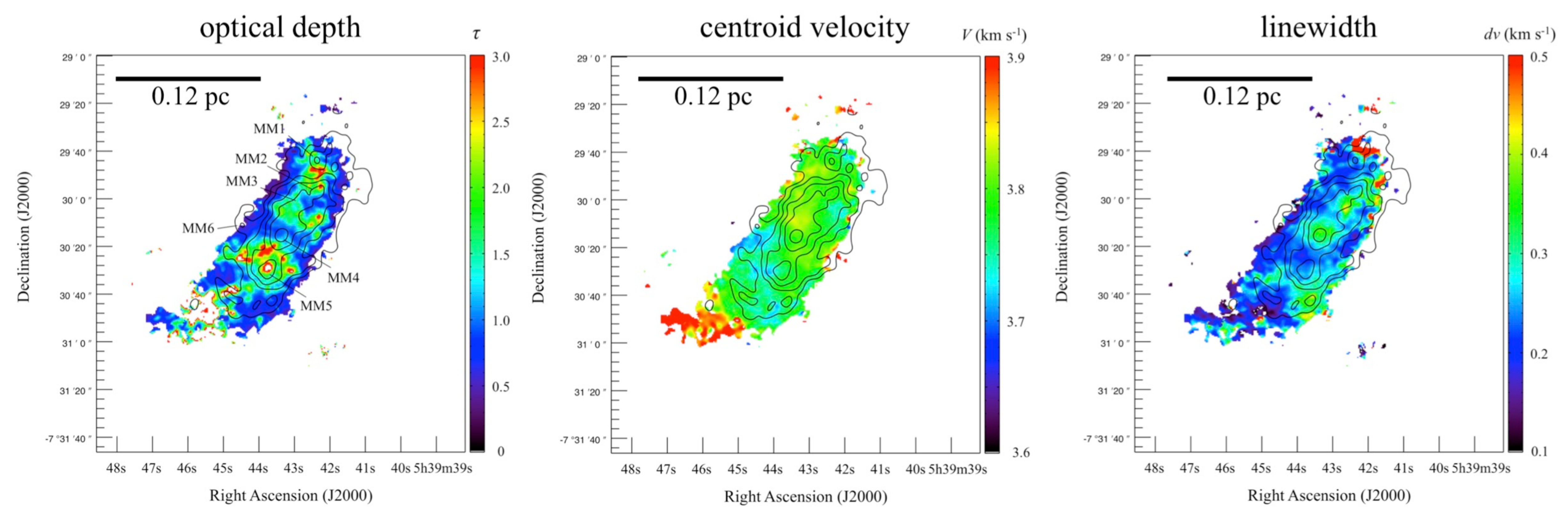}
  \end{center}
  \caption{The color maps of the optical depth ($\tau$), LSR velocity ($V_{\rm LSR}$), and linewidth ($dv$) of N$_2$H$^+$ $J=1-0$, $F_1=1-1$, $F=0-1$ emission derived from by Gaussian fitting of the weakest hyperfine component. The contours are the 3 mm dust continuum  emission from 3$\sigma$ with 2$\sigma$ steps. The fitting is performed for the pixels above 5$\sigma$.
  The optical depth is derived by assuming that the maximum intensity of N$_2$H$^+$ $J=1-0$ corresponds to the excitation temperature.}
  \label{n2h_fit}
\end{figure*}

\subsubsection{Analysis of N$_2$H$^+$ spectra}
Because N$_2$H$^+$ has hyperfine structure, we derive the optical
depth ($\tau$), LSR velocity ($V_{\rm LSR}$), linewidth ($dv$), and excitation temperature ($T_{\rm ex}$) assuming a uniform excitation temperature in the same way as \citet{oha14,oha16} using IDL fitting code.
However,  we were not able to estimate the optical depth values toward the dust peaks because N$_2$H$^+$ line profiles show some features of high optical depths.
Such extreme optical depths have been also observed in infrared dark clouds (IRDC) \citep[e.g.,][]{san13}.
Figure \ref{n2h_profile1} shows an example of N$_2$H$^+$ profile of one pixel spectra obtained by the ALMA-ACA combined image at the dust peak position.
The intrinsic intensity ratios of each hyperfine component are represented by the red segments at the bottom of Figure \ref{n2h_profile1} in the optically thin limit \citep{tin00}.
The observed intensity-profile does not follow the intrinsic intensity-profile, indicating that most of the hyperfine intensities are saturated.
Therefore, we only use the weakest and the most optically thin component ($J=1-0$, $F_1=1-1$, $F=0-1$) to derive physical parameters.
The excitation temperature $T_{\rm ex}$ is assumed to be the maximum intensities of the main component adding the cosmic microwave background of 2.7 K in each position, ranging from 4 to 6 K. Then we derive the optical depth by using the radiative transfer formula assuming the filling factor of unity. We also derive  LSR velocity, and linewidth by gaussian fitting of  the optically thinnest line.
The fitting is performed for the pixels above 5$\sigma$.
We should note that the 3 mm dust emission is optically thin ($\tau_{\rm dust}<0.1$) and would not affect the line profile.

\begin{figure}[htbp]
\setcounter{figure}{8}
  \begin{center}
  \includegraphics[width=8.5cm,bb=0 0 1523 1119]{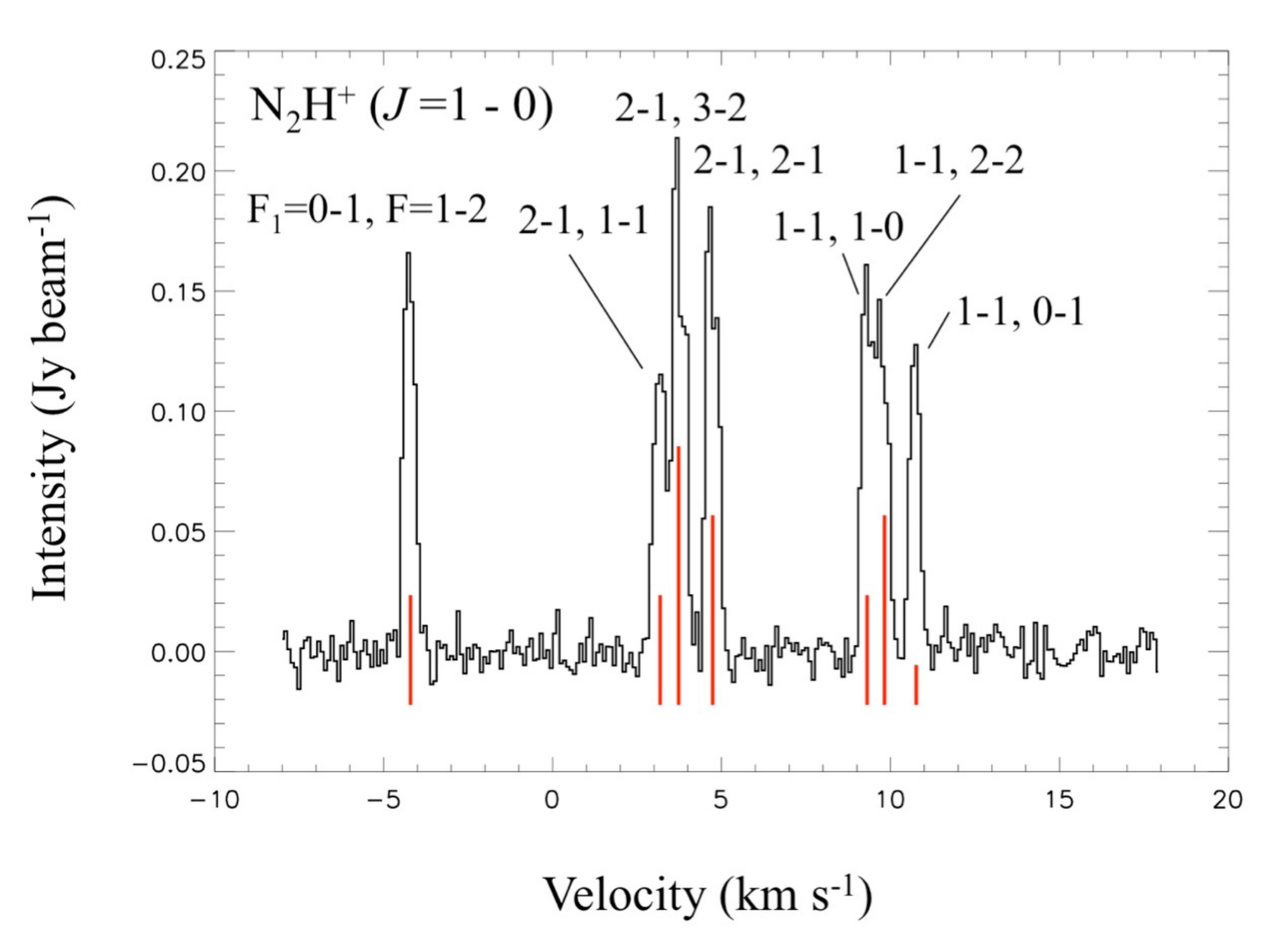}
  \end{center}
  \caption{N$_2$H$^+$ ($J=1-0$) profile at the MM4 peak position. The velocity resolution is 0.098 kms $^{-1}$.}
  \label{n2h_profile1}
\end{figure}

\begin{figure}[htbp]
  \setcounter{figure}{10}
  \begin{center}
  \includegraphics[width=8.5cm,bb=0 0 1523 1119]{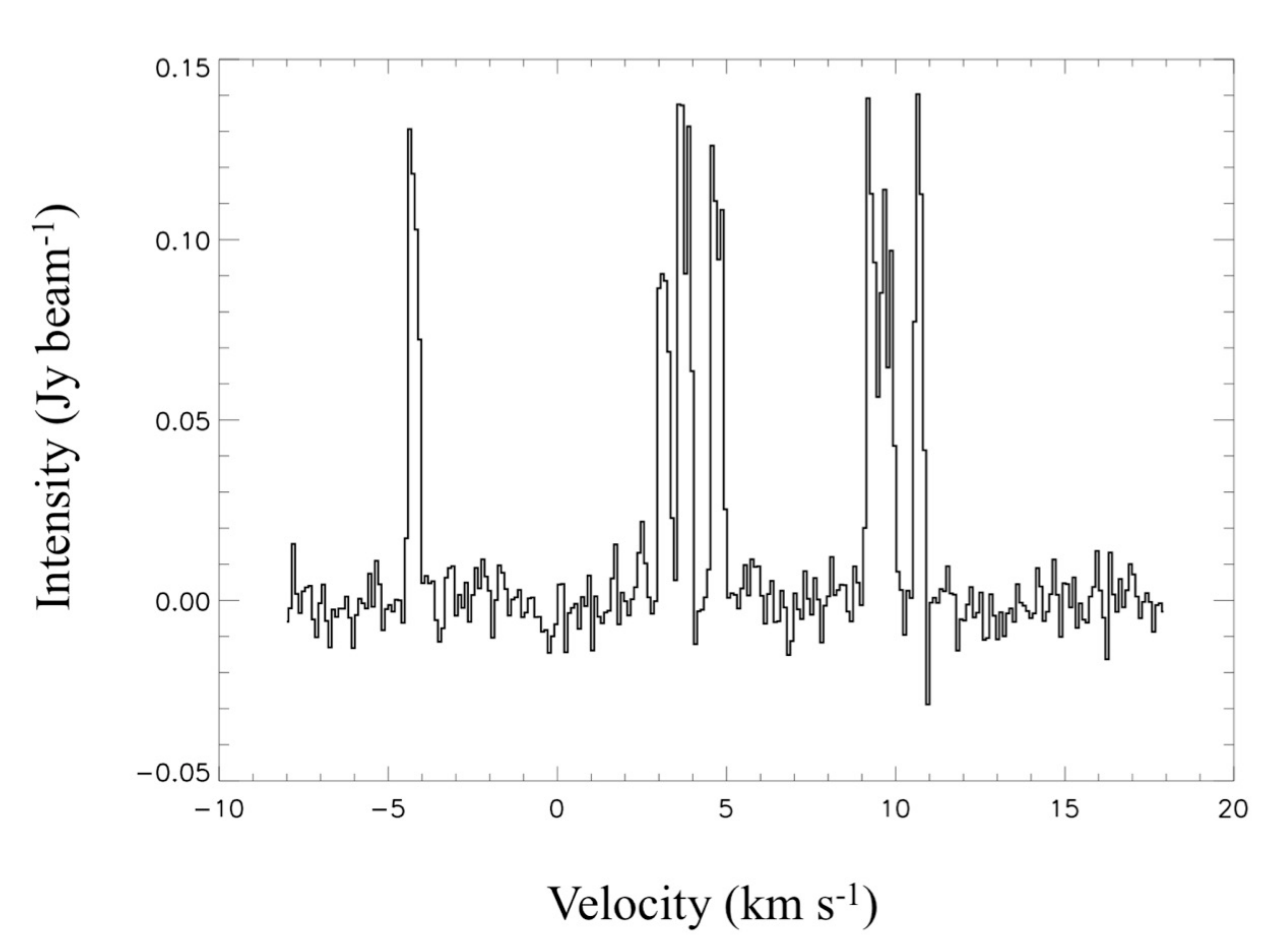}
  \end{center}
  \caption{The same as Figure \ref{n2h_profile1} but at the MM5 position.}
  \label{n2h_profile2}
\end{figure}

\begin{figure}[htbp]
  \begin{center}
  \includegraphics[width=8.5cm,bb=0 0 1523 1119]{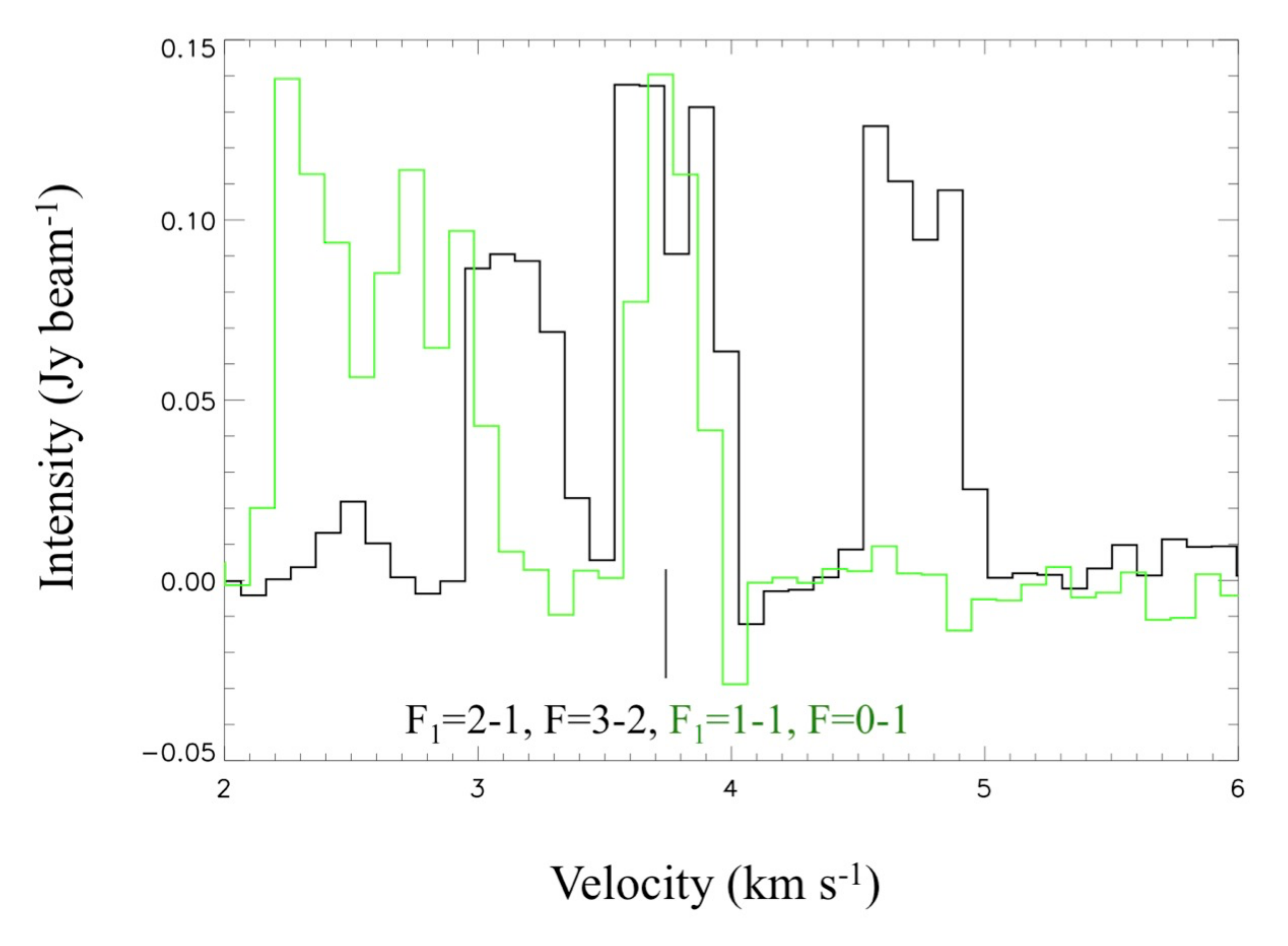}
  \end{center}
  \caption{The black line represents N$_2$H$^+$ $F_1=2-1$, $F=3-2$ profile at the MM5 position with the rest frequency of 93.173777 GHz and the green line represent the $F_1=1-1$, $F=0-1$ profile with the rest frequency of 93.171621 GHz.}
  \label{n2h_profile3}
\end{figure}

Figure \ref{n2h_fit} shows the color maps of these parameters with the 3 mm dust continuum contours (from 3$\sigma$ with 2$\sigma$ steps).
The color map of optical depth indicates that even the weakest N$_2$H$^+$ hyperfine line is moderately optically thick ($\tau\gtrsim1$) toward the condensations.
In particular, we were not able to calculate the optical depth due to the saturation of the lines in the southern dust local peak (MM5).
Figure \ref{n2h_profile2} shows the profile toward the MM5 peak position. This profile shows that the weakest hyperfine has the strongest intensity peak. 
Furthermore, we find two velocity components in the main hyperfine emission of $F_1=2-1$, $F=3-2$. 
Figure \ref{n2h_profile3} shows zoom up profiles in the same position, MM5. The black line represents the $F_1=2-2$, $F=3-2$ emission at a rest frequency of 93.173777 GHz and the green line represents the $F_1=1-1$, $F=0-1$ emission at a rest frequency of 93.171621 GHz. 
The black line shows a ``dip'' in the emission profile at 3.7 km s$^{-1}$, while the green line reachs a peak in this gap. Thus, the double peak feature seen in the strongest hyperfine component would be caused by self-absorption effect. 
In a similar case,  a self-absorption in NH$_3$ was found toward the IRDC G028.23-00.19, prestellar clump \citep{san17}. The self-absorption of N$_2$H$^+$ and NH$_3$ lines suggests that the inner part region has higher density and higher excitation temperature than surrounding N$_2$H$^+$ or NH$_3$ emission regions.

We also find that the optical depth is not correlated well with the dust continuum emission.
The MM4 is the peak position in the dust continuum but the optical depth seems to be lower than that in the surrounding region, which may suggest  the lower ionization degree in the dense part of the core or the depletion of N$_2$H$^+$  at the densest parts. 
The N-bearing molecules of N$_2$H$^+$ and NH$_3$ are suggested to be depleted  at a density of $\gtrsim10^6$ cm$^{-3}$ \citep[e.g.,][]{ber02,pag05}.

In the color map of $V_{\rm LSR}$, there is no evident velocity variation along the filament.
Systematic velocity fluctuations is also not found along the ridge of the continuum image.
This could be due to the impossibility of tracing velocity fields using optically thick lines (most regions have $\tau\gtrsim1$) or the insufficient velocity resolution (0.098 km s$^{-1}$) of our observation to identify infall or fragment motions.
On the other hand, in the linewidth map, we find that the linewidth increases toward the condensations. Because the central condensation does not have a high optical depth  ($\tau\sim1$) in comparison with the surrounding part, broad linewidth is not due to the saturation effect of the line but indicate motions (infall, expansion, or turbulent) toward the condensation. 
Considering that the N$_2$H$^+$ molecule is partially depleted and the linewidth becomes broader toward the MM4, this condensation may be beginning to collapse.  However, our current velocity resolution and the molecular lines are not enough to resolve and investigate the motions in details.
It is highly needed to investigate the nature of the enhanced linewidth, i.e, the presence of  blue-skewed profile suggesting infall, with higher spectral resolution.

\begin{figure*}[htbp]
  \begin{center}
  \includegraphics[width=19cm,bb=0 0 2943 977]{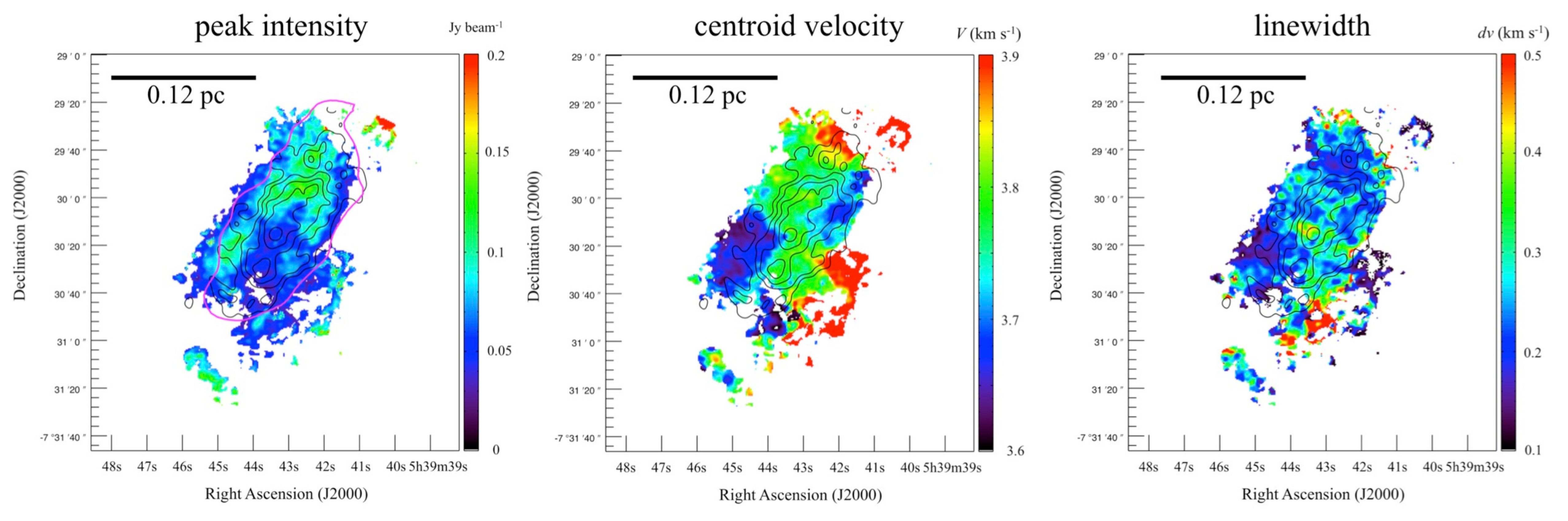}
  \end{center}
  \caption{The color maps of the peak intensity (Jy beam$^{-1}$), LSR velocity ($V_{\rm LSR}$), and linewidth ($dv$) of CH$_3$OH ($J=2_{0,2}$ $_-$ $1_{0,1}$ A$^{+}$)  emission derived by the Gaussian fitting. The fitting is performed for the pixels above 5$\sigma$ of the CH$_3$OH emission. Black contours show the ALMA-ACA 3 mm dust continuum emission from 3$\sigma$ with 2$\sigma$ steps. The magenta contour shows the 5$\sigma$ contour of the ACA 3 mm dust image for averaging the CH$_3$OH profile.}
  \label{ch3oh_fit}
\end{figure*}

Even though N$_2$H$^+$ is highly optically thick, we roughly estimate the N$_2$H$^+$ abundance at the dust peak position.
In the dust peak, we assume the excitation temperature of $T_{\rm ex}=5.8$ K from the maximum intensity. The optical depth of the $F_1=1-1$, $F=0-1$ line is derived to be $\tau=0.9\pm0.1$.
Then, the column density can be calculated by following \citet{san12,man15}.  We derive $N_{\rm N_2H^+}=(3.4\pm0.4)\times10^{13}$ cm$^{-2}$.
On the other hand, the 3 mm dust continuum shows $N({\rm H_2})=9.9\times10^{22}$ cm$^{-2}$.
Therefore, the N$_2$H$^+$ abundance is derived to be $X({\rm N_2H^+})=(3.4\pm0.4)\times10^{-10}$, which is similar to other low-mass star forming regions \citep[e.g.,][]{cas02,taf02,taf06,ket04,fri10}.

\subsubsection{Analysis of CH$_3$OH spectra}

The CH$_3$OH transitions observed are ($2_{-1}-1_{-1}$ $E$), ($2_0-1_0$ $A^+$), and ($2_0-1_0$ $E$).
Assuming LTE and optically thin conditions, we can estimate the column density ($N_{\rm CH_3OH}$) and rotation temperature ($T_{\rm rot}$) using the rotation diagram technique \citep[e.g.,][]{bla87}.
However, CH$_3$OH ($2_0-1_0$ $E$) has a low S/N ratio. In order to have a significant detection in all 3 lines, we average the profiles from the whole region inside the 5$\sigma$ contour of the ACA 3 mm dust continuum image. The averaging region shows the magenta contour in Figure \ref{ch3oh_fit}.
Following the population diagram methods \citep[e.g.,][]{gol99}, we fit the three CH$_3$OH lines. 
We follow the procedure described in \citet{san13}, which is optimized for cold gas.
The obtained column density and rotation temperature are $N_{\rm CH_3OH}=(1.1\pm0.1)\times10^{13}$ cm$^{-2}$ and $T_{\rm rot}=10.8\pm0.4$ K.
This rotation temperature is consistent with $T_{\rm rot}=10.6$ K derived by using the NH$_3$ ($J,K=1,1$) and ($J,K=2,2$) lines \citep{oha16}.
The rotation temperature is the same value even if the different molecular lines are used, suggesting the isothermal core with the LTE condition.
The optical depth of the strongest transition  ($2_0-1_0$ $A^+$) is $\sim0.1$, and hence, it is optically thin.
The average H$_2$ column density is derived to be $N({\rm H_2})=3.8\times10^{22}$ cm$^{-2}$ from the 3 mm dust continuum observations.
Therefore, the CH$_3$OH abundance is  $X({\rm CH_3OH})\sim2.9\times10^{-10}$.
\citet{som15} estimated the abundance of CH$_3$OH to be $\sim10^{-9}$ toward the cold quiescent core, TMC-1 (CP).
\citet{taf06} also estimated the abundance to be $X({\rm CH_3OH})\sim6\times10^{-10}$ in the L1498 and L1517B prestellar cores.
TUKH122 has similar CH$_3$OH abundance to these prestellar cores.
In particular, the CH$_3$OH abundance can be even lower at the central part because this value is the average over the whole region.
Assuming that the CH$_3$OH ($2_0-1_0$ $A^+$) line is optically thin and $T_{\rm rot}=10.8$ K, we estimate the column density, $N_{\rm CH_3OH}$, at the dust peak position MM4 by using the rotation temperature derived above. 
By comparing the velocity-integrated intensities, the  column density is derived to be $N_{\rm CH_3OH}=1.3\times10^{13}$ cm$^{-2}$ at this dust peak position, and then the CH$_3$OH abundance is  $X({\rm CH_3OH})\sim1.3\times10^{-10}$.

Figure \ref{ch3oh_fit} shows in color maps of the peak intensity, $V_{\rm LSR}$, and linewidth of CH$_3$OH.
The peak temperature map clearly shows the shell like structure and the dust condensations are located within this shell.
The $V_{\rm LSR}$ map shows some systematic variations within the core. The northern edge is slightly red-shifted in velocity, while the southern edge is blue-shifted, suggesting core rotation or gas inflow motions along the filamentary structure. Furthermore, the south-eastern side is blue-shifted velocity, while the shouth-western side is red shifted.
The velocity structure is complicated but the velocity difference might indicate the contraction or twisted motions of the core.

The linewdith map shows no systemic variations along the filament. However, the dust peak position (MM4) has  a broad linewidth of $\Delta v\sim0.4$ km s$^{-1}$, which is the same tendency with that in the N$_2$H$^+$. These broad linewidths would thus suggest infall motions.
Interestingly, the linewidth color maps of N$_2$H$^+$ and CH$_3$OH show a small value of $dv\lesssim0.2$ km s$^{-1}$ at the edge of the condensations and our velocity resolution marginally resolve the lines. The linewidth of 0.2 km s$^{-1}$ is close to the thermal motion because thermal linewidth corresponds to be 0.14 km s$^{-1}$ for N$_2$H$^+$ and 0.13 km s$^{-1}$ for CH$_3$OH at 12 K. 
The narrow linewith of the core edge indicates that the turbulence is almost dissipated in the filament, and the non-thermal motions within the core could indicate infall motions or some systemic velocity variations such as rotation and fragmentation.

The important difference between TUKH122 and the other prestellar cores in nearby clouds is that TUKH122 consists of at least three dense condensations within the one CH$_3$OH shell.
The TUKH122 core is more massive than low-mass dense cores ($\sim$ $M_\odot$) \citep[e.g.,][]{oni02,alv07} even though the size is comparable, implying that the TUKH122 core is denser. The denser core would enhance CH$_3$OH shell structure because UV radiation is only irradiated on the outer layer of the core.

\section{Discussion}

\subsection{Density profile}

\begin{figure}[htbp]
  \begin{center}
  \includegraphics[width=8.5cm,bb=0 0 1912 1097]{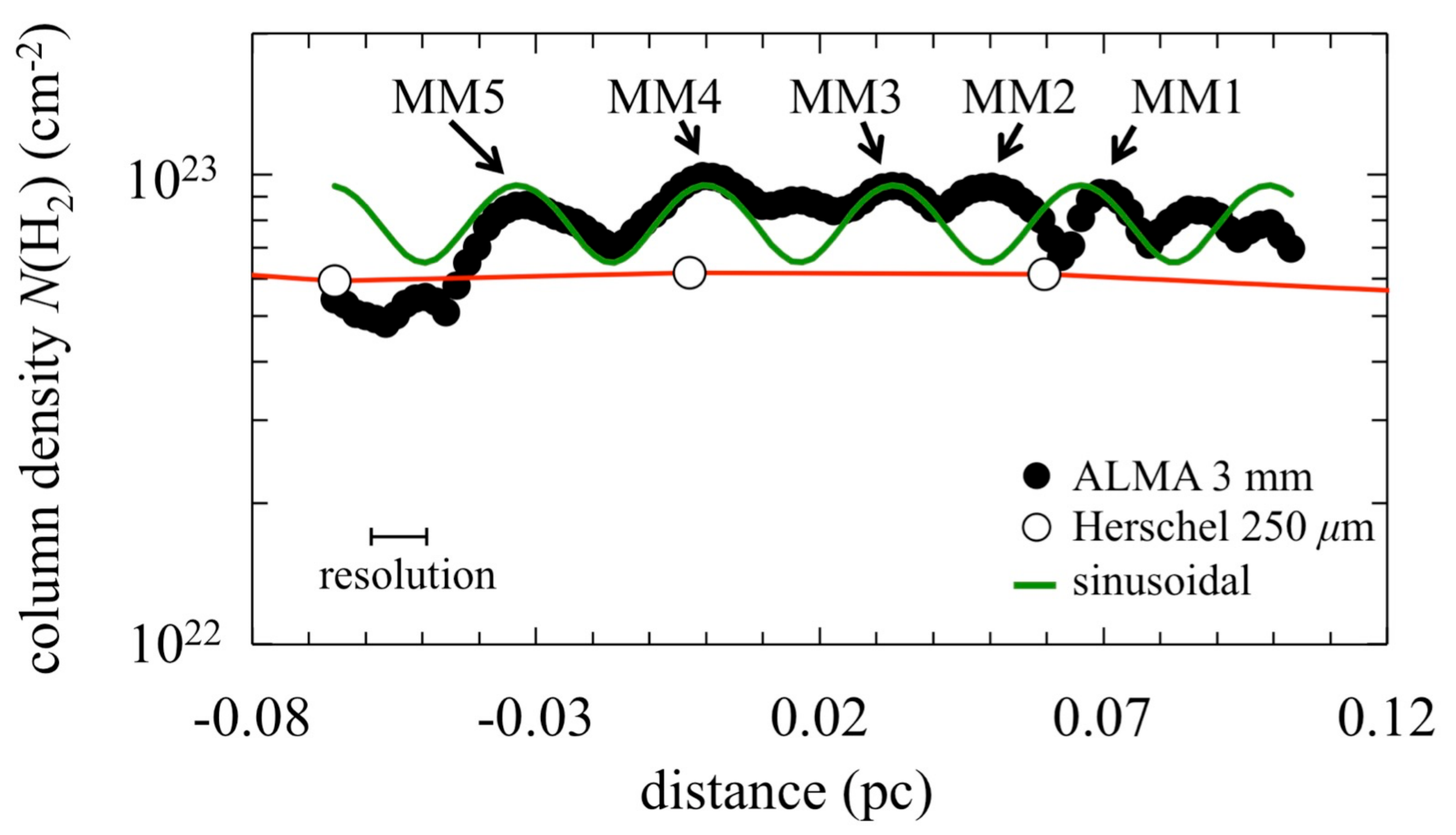}
  \end{center}
  \caption{The maximum column density profile along the elongated core taken by ALMA-ACA observations and the {\it Herschel} 250 $\mu$m observations. The spatial resolution is $\sim0.01$ pc. 
  The green line represents a sinusoidal motion with an interval of 0.035 pc. 
  }
  \label{along}
\end{figure}

To investigate the formation of the condensations, we  study the density structure along the filament.
Figure \ref{along} shows the column density profiles taken from the maximum position along the filament direction (see also Figure \ref{cont_combine_pa}).
The reference position here is the MM4 condensation corresponding to the peak flux in the ALMA-ACA observations.
The {\it Herschel} single dish observations indicate almost flat density structure due to the large beamsize.
On the other hand, the ALMA-ACA observations show column density variations.
The green line represents a sinusoidal column-density variation with an interval of 0.035 pc to visually match the observed variations.
The southern part (MM3, 4, and 5) seems to be nicely fitted by this sinusoidal variation.

The interval of 0.035 pc is consistent with the separations between neighbor sources in the OMC-2 and 3 regions \citep{tak13,kai17}.
In OMC-1, another part of the Orion A cloud, a separation of $\sim0.01$ pc has been reported  by \citet{tei16,pal17}.
Since the OMC-1 is the most massive part in the Orion A cloud, the fragmentation might occur at denser region and the separations would then become shorter.
The separations and fragmentation processes have also been studied in other regions including high mass star forming regions and suggested to be governed by thermal Jeans processes \citep[e.g.,][]{kai13,beu15,pal15,bus16}.
Assuming that the fragmentation occurs by the Jeans instability, we can estimate the separation of the fragment at this region.
With the assumption of an infinite size and a uniform density, the Jeans length is described as follows \citep{jea02}:
\begin{equation}
\lambda_{\rm Jeans}=\sqrt{\frac{\pi c_s^2}{G\rho_0},}
\label{jeans}
\end{equation}
where $G$ is the gravitational constant, $\rho_0$ is the mean density, and $c_s$ is the sound speed of 0.2 km s$^{-1}$ at 12 K.
\citet{tak13} also analyzed the Jeans length in the case of an infinitely long static and cylindrical isothermal cloud following \citet{nak93,wis98}
 and described as
\begin{equation}
\lambda_{\rm Jeans}\sim\frac{20 c_s}{\sqrt{4\pi G\rho_0}}.
\label{jeans2}
\end{equation}
From the column density profile of the ALMA-ACA observations in Figure \ref{perpen}, we derive the central density, $\rho_{c}\sim5.5\times10^5$ cm$^{-3}$ assuming the numerical factor of $A=\pi/2$.
The Jeans length is derived to be $\sim0.031$ pc and $\sim0.099$ pc, respectively following the equations of (\ref{jeans}) and (\ref{jeans2}) assuming the temperature of 12 K.
The thermal Jeans instability with the assumption of an infinite and a uniform density, the equation (\ref{jeans}), would be better explained to form the condensations.
Therefore, the fragmentation occurs within the elongated core having a density of $\sim10^5$ cm$^{-3}$ and the denser condensations with a density of $\sim10^{6-7}$ cm$^{-3}$ are formed by the thermal Jeans instability. 
Note that assuming $\lambda_{\rm Jeans}=0.035$ pc, the density is derived to be $n({\rm H_2})\sim4.4\times10^{5}$ cm$^{-3}$ from the equation (\ref{jeans}).

The northern part (MM1 and 2) seems to have a shorter interval of $\sim0.02$ pc.
An inclination might cause a shorter interval even if these condensations have the same interval of $\sim0.035$ pc or the Jeans instability at higher densities may occur. 
We should also note that theses northern condensations are located on the CH$_3$OH shell structure, suggesting the edge of the TUKH122 core.
Therefore, the edge instabilities may also affect the separation of the condensations \citep{pon11,pon12}.
The lower mass condensations with shorter intervals may be formed in the edge region.

We also discuss a possibility whether the condensations will merge or not.
From the CH$_3$OH observations, the relative velocity of the surrounding condensation for MM 4 are $0.02-0.05$ km s$^{-1}$.
To merge the condensations, we need at least $4\times10^5$ yr assuming the separation of 0.02 pc and velocity of 0.05 km s$^{-1}$.
On the other hand, the free fall time at a density of $10^6$ cm$^{-3}$ will be $4\times10^4$ yr.
Therefore, the star formation can be started before these condensations merge.
It is unlikely that these condensations merge to a larger scale.
However, the magnetic field may be supporting the condensations and makes the star formation time scale longer \citep[e.g.,][]{cru04,war07}.
It is suggested that the ambipolar diffusion timescale is about 10 times longer than the free fall time \citep{nak98}.
In this case, the merging might be possible.

\subsection{Possibility of multiple star formation}
In the previous section, we identified the 6 condensations with a mass of $\sim0.1-0.4$ $M_\odot$ in the core and they are gravitationally bound.
However, the total mass of the condensations is only 1.2 $M_\odot$ and a large portion of the mass is located outside of them.
One possibility of the small mass fraction of the condensations is that we may only detect the peak region of the condensations, and the condensations may have larger radius and mass.
For example, if the condensations have unstable Bonnor-Ebert sphere, they have a large contrast in density between the center and outer part.
The condensations might be able to identify only the center part due to our spatial resolutions and sensitivities.
Actually, the size of the condensations is derived to be $r\sim0.006$ pc even though we find the separation of $0.02-0.035$ pc between them.
If we use the separations as the size of these condensations, the mass is derived to be $0.3-1$ $M_\odot$.
Even if the condensations have masses of $0.3-1$ $M_\odot$, they only have a small amount of the mass and the virial parameters are still lower than unity. These condensations may collapse immediately.
Another possibility is that  very low-mass stars ($\sim0.03-0.1$ $M_\odot$) are formed in these condensations in this region.
In either case, the multiple star system may be formed in the  core along the filament.
We suggest that the fragmentation is still important in the dense core to form multiple stars.

\section{conclusions}

We have presented the ALMA high spatial resolution observations of 3 mm dust continuum, the hyperfine-components lines of N$_2$H$^+$ ($J=1-0$), and the CH$_3$OH ($J_K=2_K-1_K$) lines toward the TUKH122 core.
This core is likely on the verge of star formation because the turbulence is almost completely dissipated and is a chemically evolved core among starless core in the Orion A cloud.
Our main results are summarized as follows.

1.  The column density profile perpendicular to the parent filamentary structure including the elongated core is  similar to the Plummer-like function. The {\it Herschel} observations show the power law of $p=1.6$, while the ALMA-ACA observations shows a power law of $p=4.0$. This difference will be explained in terms of the missing flux of the interferometric observations.
The condensations are embedded in the parent elongated core.

2. ALMA 12-m dust continuum emission shows  compact dense condensations, while ACA 7-m image shows the oval structure due to the larger beamsize and recovering the extended emission.
By combining these data and applying the dendrogram method, we find condensations within the core with a mass range of $0.1-0.4$ $M_\odot$ and a deconvolved radius of $0.003-0.01$ pc along the filament.
The densities are $n\sim10^{6-7}$ cm$^{-3}$, which are an order of magnitude higher than typical low-mass prestellar cores.

3.  We find different distributions between N$_2$H$^+$ and CH$_3$OH emission. The N$_2$H$^+$ distribution is similar to the dust continuum but seems to be frozen-out in the dust peak MM4, while the CH$_3$OH distribution shows a shell-like structure. The extended CH$_3$OH distribution suggests that the CH$_3$OH molecule formed on dust grain is released into gas phase by non-thermal desorption such as desorption caused by cosmic-ray induced UV radiation. We also find the absorption feature in the N$_2$H$^+$ $F_1=2-1$ $F=3-2$ component toward the condensation, MM5, due to the self-absorption effect.
Even though N$_2$H$^+$ has been used as an optically thin dense gas tracer, we show that this molecule becomes optically thick in this core.

4.  The separations of these condensations are  $\sim0.035$ pc in the southern part, which is similar to that in the OMC-2/3 region, the active star forming region in the Orion A cloud. 
The Jeans length corresponds to be $n({\rm H_2})\sim4.4\times10^{5}$ cm$^{-3}$, which is almost consistent with that of the central density of the core.  Therefore the fragmentation still occurs in the prestellar core by thermal Jeans instability at a density of $\sim10^5$ cm$^{-3}$ and the denser condensations with a density of $10^{6-7}$ cm$^{-3}$ are formed.
In the northern part corresponding to the CH$_3$OH shell region, we find that the condensations are less massive with shorter intervals, which may imply that the edge instabilities affect the fragmentation.

5. The condensations are not in virial equilibrium and may collapse immediately unless the magnetic field counterbalances gravity. However, the total mass of the condensations is a small fraction of the total TUK122 core mass.
The condensations might identify only the central part due to our spatial resolutions and sensitivities and they may have masses of $0.3-1$ $M_\odot$. Another possibility is that the very low-mass stars ($\sim0.03-0.1$ $M_\odot$) are formed in the small condensations in this region.
In either case, multiple stars may be formed in the core. We suggest that the fragment due to the thermal instability is still important to form a multiple star system.

\acknowledgments
We gratefully appreciate the comments from the anonymous referee that significantly improved this article.
S.O. also thank  Kazuya Saigo for helpful discussions.
This paper makes use of the following ALMA data: ADS/JAO.ALMA\#2015.1.01025.S. ALMA 
is a partnership of ESO (representing its member states), NSF (USA) and NINS (Japan), 
together with NRC (Canada), NSC and ASIAA (Taiwan), and KASI (Republic of Korea), in 
cooperation with the Republic of Chile. The Joint ALMA Observatory is operated by 
ESO, AUI/NRAO and NAOJ.

This research has also made use of data from the Herschel Gould Belt survey (HGBS) project (http://gouldbelt-herschel.cea.fr). The HGBS is a Herschel Key Programme jointly carried out by SPIRE Specialist Astronomy Group 3 (SAG 3), scientists of several institutes in the PACS Consortium (CEA Saclay, INAF-IFSI Rome and INAF-Arcetri, KU Leuven, MPIA Heidelberg), and scientists of the Herschel Science Center (HSC).

Data analysis was in part carried out on common use data analysis computer system at the Astronomy Data Center, ADC, of the National Astronomical Observatory of Japan.



{\it Facilities:} \facility{ALMA, Herschel}

\end{document}